\pdfoutput=1
\documentclass[]{emulateapj}
\usepackage{apjfonts}
\begin{document}
\journalinfo{\today}
\slugcomment{}
\title{High-Angular-Resolution and High-Sensitivity Science Enabled by
  Beamformed ALMA}
\author{Vincent Fish\altaffilmark{1}, Walter Alef, James Anderson,
  Keiichi Asada\altaffilmark{1}, Alain Baudry, Avery
  Broderick\altaffilmark{1}, Chris Carilli\altaffilmark{1}, Francisco
  Colomer, John Conway\altaffilmark{1}, Jason Dexter, Sheperd
  Doeleman\altaffilmark{1}, Ralph Eatough, Heino
  Falcke\altaffilmark{1}, S\'andor Frey, Krisztina Gab\'anyi, Roberto
  G\'{a}lvan-Madrid, Charles Gammie, Marcello Giroletti, Ciriaco
  Goddi, Jose L.\ G\'{o}mez, Kazuhiro Hada, Michael Hecht, Mareki
  Honma\altaffilmark{1}, Elizabeth Humphreys, Violette Impellizzeri,
  Tim Johannsen, Svetlana Jorstad, Motoki Kino, Elmar K\"{o}rding,
  Michael Kramer, Thomas Krichbaum\altaffilmark{1}, Nadia
  Kudryavtseva, Robert Laing\altaffilmark{1}, Joseph Lazio, Abraham
  Loeb, Ru-Sen Lu, Thomas Maccarone, Alan Marscher\altaffilmark{1},
  Iv\'{a}n Mart\'{\i}-Vidal, Carlos Martins, Lynn Matthews, Karl
  Menten, Jon Miller, James Miller-Jones, F\'{e}lix Mirabel, Sebastien
  Muller, Hiroshi Nagai, Neil Nagar\altaffilmark{1}, Masanori
  Nakamura, Zsolt Paragi, Nicolas Pradel, Dimitrios
  Psaltis\altaffilmark{1}, Scott Ransom, Luis Rodr\'{\i}guez, Helge
  Rottmann, Anthony Rushton, Zhi-Qiang Shen\altaffilmark{1}, David
  Smith, Benjamin Stappers, Rohta Takahashi\altaffilmark{1}, Andrea
  Tarchi, Remo Tilanus, Joris Verbiest, Wouter Vlemmings, R.\ Craig
  Walker, John Wardle, Kaj Wiik, Erik Zackrisson, \& J.\ Anton Zensus}
\email{vfish@haystack.mit.edu}
\altaffiltext{1}{Editor}
\begin{abstract}
An international consortium is presently constructing a beamformer for
the Atacama Large Millimeter/submillimeter Array (ALMA) in Chile that
will be available as a facility instrument.  The beamformer will
aggregate the entire collecting area of the array into a single, very
large aperture.  The extraordinary sensitivity of phased ALMA,
combined with the extremely fine angular resolution available on
baselines to the Northern Hemisphere, will enable transformational new
very long baseline interferometry (VLBI) observations in Bands 6 and 7
(1.3 and 0.8~mm) and provide substantial improvements to existing VLBI
arrays in Bands 1 and 3 (7 and 3~mm).  The ALMA beamformer will have
impact on a variety of scientific topics, including accretion and
outflow processes around black holes in active galactic nuclei (AGN),
tests of general relativity near black holes, jet launch and
collimation from AGN and microquasars, pulsar and magnetar emission
processes, the chemical history of the universe and the evolution of
fundamental constants across cosmic time, maser science, and
astrometry.
\end{abstract}

\section{Preamble}

The science case for the ALMA beamformer reflects the broad scientific
goals of the VLBI community.  Astronomers were solicited for input at
a variety of meetings including the 10th European VLBI Network (EVN)
Symposium (Manchester 2010), the XXX URSI General Assembly (Istanbul
2011), Bringing Black Holes Into Focus: The Event Horizon Telescope
(Tucson 2012), Outflows, Winds and Jets: From Young Stars to
Supermassive Black Holes (Charlottesville 2012), From Stars to Black
Holes: mm-VLBI with ALMA and Other Telescopes (Garching 2012), and the
11th EVN Symposium (Bordeaux 2012).  This document includes input from
scientists in all ALMA regions and is designed to be a living
document, envisaged to evolve as new science areas are developed and
existing areas refined by the community.

The case for science with beamformed ALMA is broad, ranging from
Galactic to extragalactic science and touching on questions of vital
importance for fundamental physics.
\begin{itemize}
\item ALMA VLBI observations of nearby supermassive black holes in
  Bands 6 and 7 will produce the first Schwarzschild-radius-scale
  images of nuclear black hole accretion disks and jets.  With this
  resolution, astronomers will be able to perform new tests of general
  relativity and the no-hair theorem in a strong-field environment and
  understand in detail how magnetized plasmas are accreted and
  launched into relativistic jets that extend out to kiloparsec scales
  (Section~\ref{blackholes}).
\item High-resolution imaging of AGN jets in conjunction with VLBI
  arrays in Bands 3 and 1 will clarify the internal jet structure, the
  role of magnetic fields in jet launch and collimation, and
  connections with very-high-energy (VHE) photon emission
  (Section~\ref{agn}).
\item Observations of pulsars at millimeter wavelengths will shed
  light on the processes that produce coherent centimeter-wavelength
  emission, and searches for pulsars toward the Galactic Center, which
  can be done commensally with other Galactic Center observations.
  The detection of a pulsar in orbit near Sgr~A* that could also be
  used to probe the Kerr metric around the black hole with high
  precision (Section~\ref{pulsars}).
\item Spectral-line VLBI of absorbing systems will measure the
  chemical evolution of the universe over cosmic time and test whether
  the fundamental constants of nature are actually variable
  (Section~\ref{absorbers}).
\item VLBI observations of masers will refine estimates of the
  physical conditions and dynamics in the circumstellar gas around
  young stellar objects as well as in the circumnuclear environment of
  AGN (Section~\ref{masers}).
\item Astrometry will clarify the structure of the Milky Way and
  obtain geometric distances to Galactic and extragalactic objects
  (Section~\ref{astrometry}).
\end{itemize}
In order to address these scientific topics, phased ALMA will
primarily be used as a crucial station in millimeter and submillimeter
VLBI arrays.  However, it will also be valuable as a stand-alone
instrument (e.g., for high-frequency pulsar observations).

\section{Enhanced VLBI Capabilities Provided by the ALMA Beamformer
  System} 

The ALMA beamformer will most frequently be used in conjunction with
other VLBI telescopes.  ALMA's most significant contributions to VLBI
arrays will be superlative sensitivity and higher angular resolution.
The aggregated collecting area of 50 ALMA 12-m telescopes is
equivalent to that of a single telescope with a diameter of $\sim
84$~m.  This large aperture, combined with excellent typical
atmospheric conditions and low receiver noise temperatures, means that
phased ALMA will be the most sensitive element in VLBI arrays.

The expected noise characteristics of ALMA in Bands 3, 6, 7, and 9 are
summarized in Table~\ref{tab-noise}.  In Band 3 the system equivalent
flux density (SEFD) of ALMA will be a factor of two better than the
Green Bank Telescope (GBT) and nearly two orders of magnitude better
than telescopes in the Very Long Baseline Array (VLBA).  Band 3
continuum observations between ALMA and the GBT can provide a
single-polarization rms noise of less than 20~$\mu$Jy in 1~hr in a
bandwidth of 4~GHz.  Spectral line observations (as, for instance, of
86~GHz SiO masers) will achieve an rms of nearly 2~mJy in a
1~km\,s$^{-1}$ velocity channel.  At observing frequencies of 230 and
345~GHz, phased ALMA will improve upon the sensitivity of the phased
SMA by more than a factor of 7 in rms noise on corresponding
baselines.

\begin{deluxetable}{cccccc}
\tablewidth{\hsize}
\tablecaption{Expected noise characteristics of phased ALMA\label{tab-noise}}
\tablehead{
  \colhead{Band} &
  \colhead{Frequency} &
  \colhead{Weather} &
  \colhead{$T_\mathrm{sys}$} &
  \colhead{Aperture} &
  \colhead{SEFD} \\
  \colhead{Number} &
  \colhead{(GHz)} &
  \colhead{(octile)} &
  \colhead{(K)} &
  \colhead{Efficiency} &
  \colhead{(Jy)}
}
\startdata
3 & \phantom{0}95 & 4th & \phantom{00}70 & 0.71 & \phantom{00}48 \\
6 &           230 & 4th & \phantom{00}97 & 0.68 & \phantom{00}70 \\
7 &           340 & 4th & \phantom{0}180 & 0.63 & \phantom{0}140 \\
9 &           675 & 2nd &           1150 & 0.43 &           1310 \\
9 &           675 & 1st & \phantom{0}850 & 0.43 & \phantom{0}970
\enddata
\tablecomments{System temperatures are taken from the ALMA Sensitivity
  Calculator.  Aperture efficiencies are from Table 8.3 of the ALMA
  Technical Handbook (2012-06-04).  Phasing losses have not been
  included.  Median weather conditions are assumed at 340~GHz and
  below since VLBI scheduling will be constrained by conditions at
  other sites.}
\end{deluxetable}

\begin{figure}
\resizebox{0.49\hsize}{!}{\includegraphics{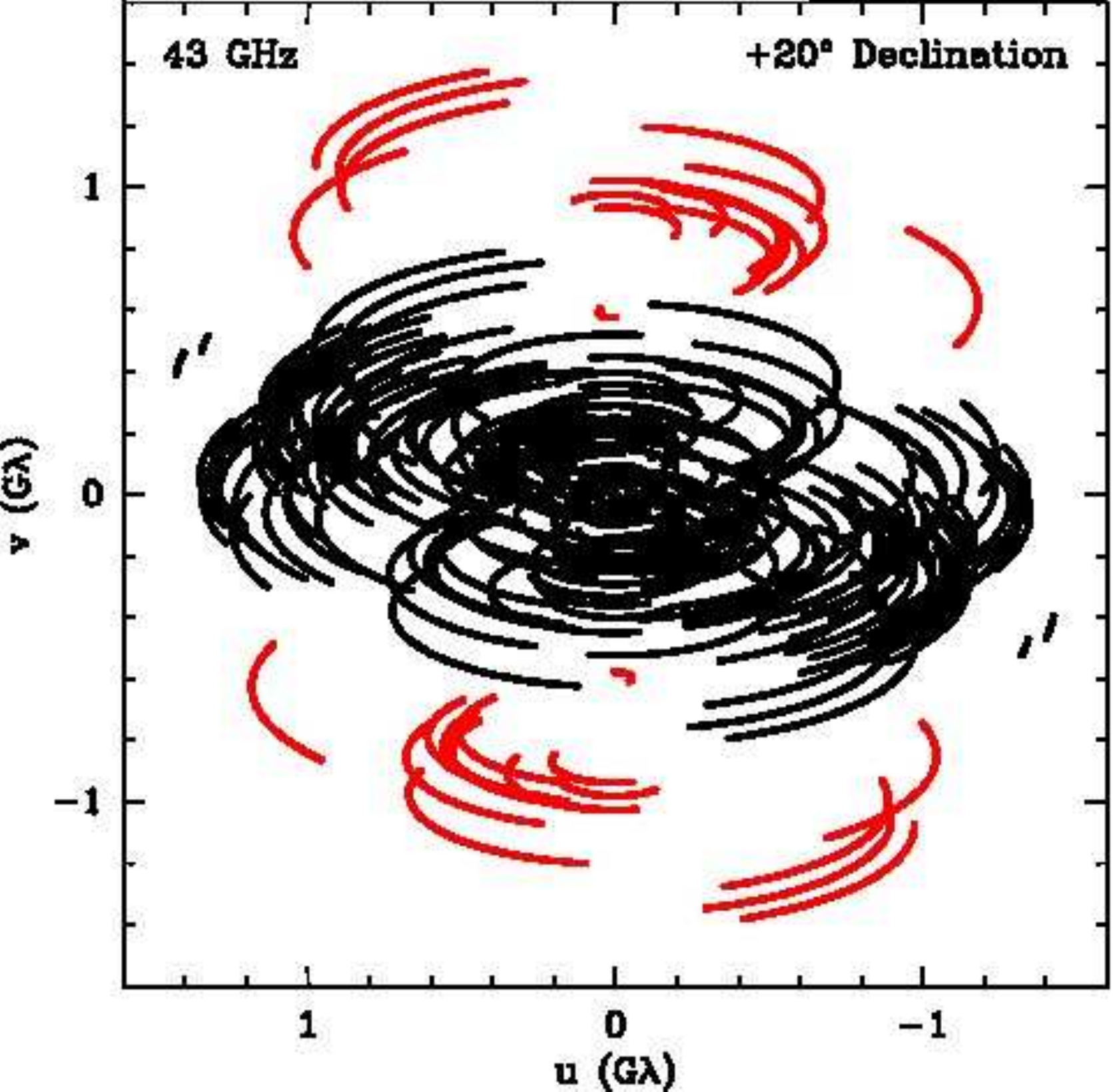}}
\resizebox{0.49\hsize}{!}{\includegraphics{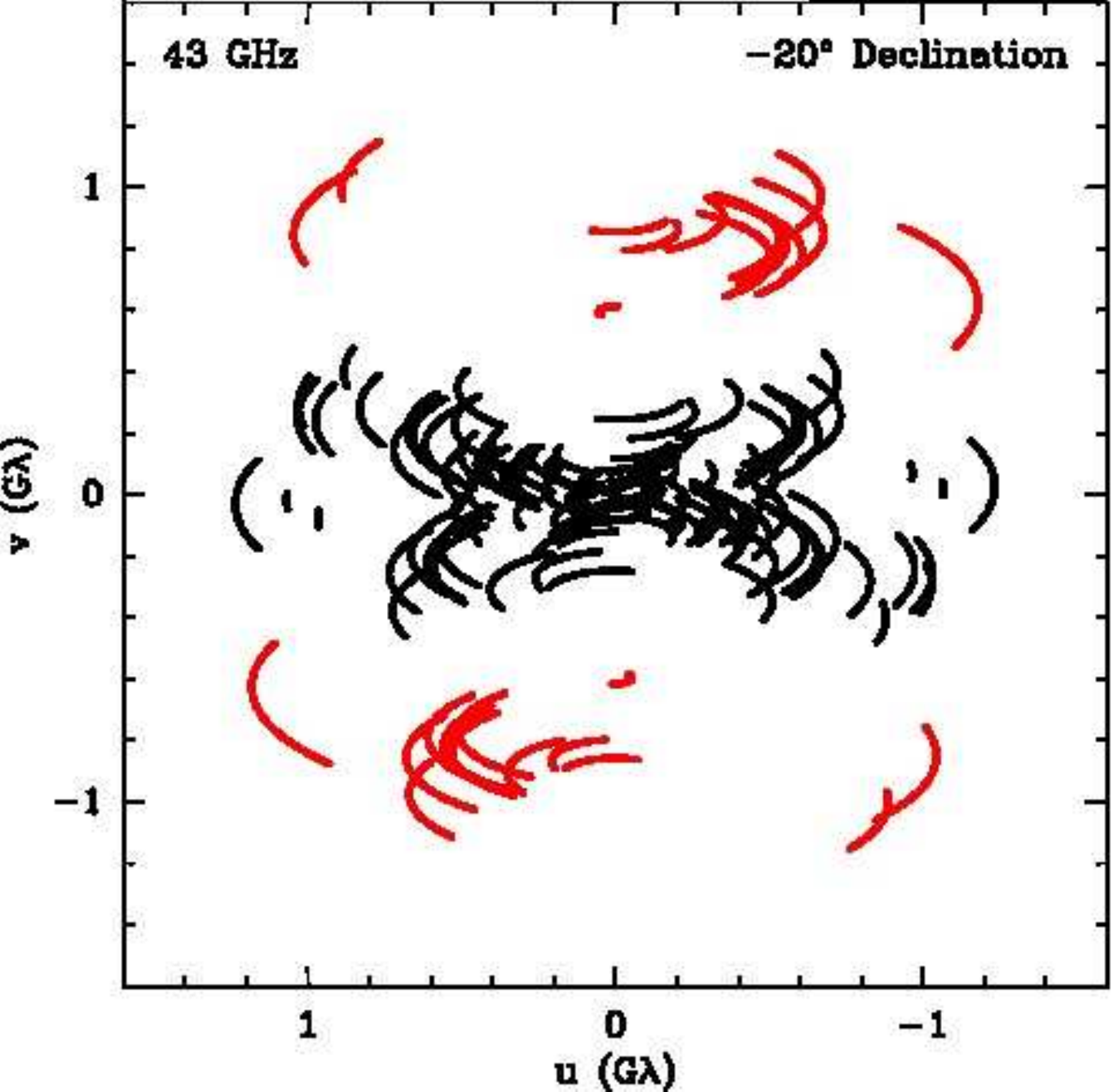}}\\
\resizebox{0.49\hsize}{!}{\includegraphics{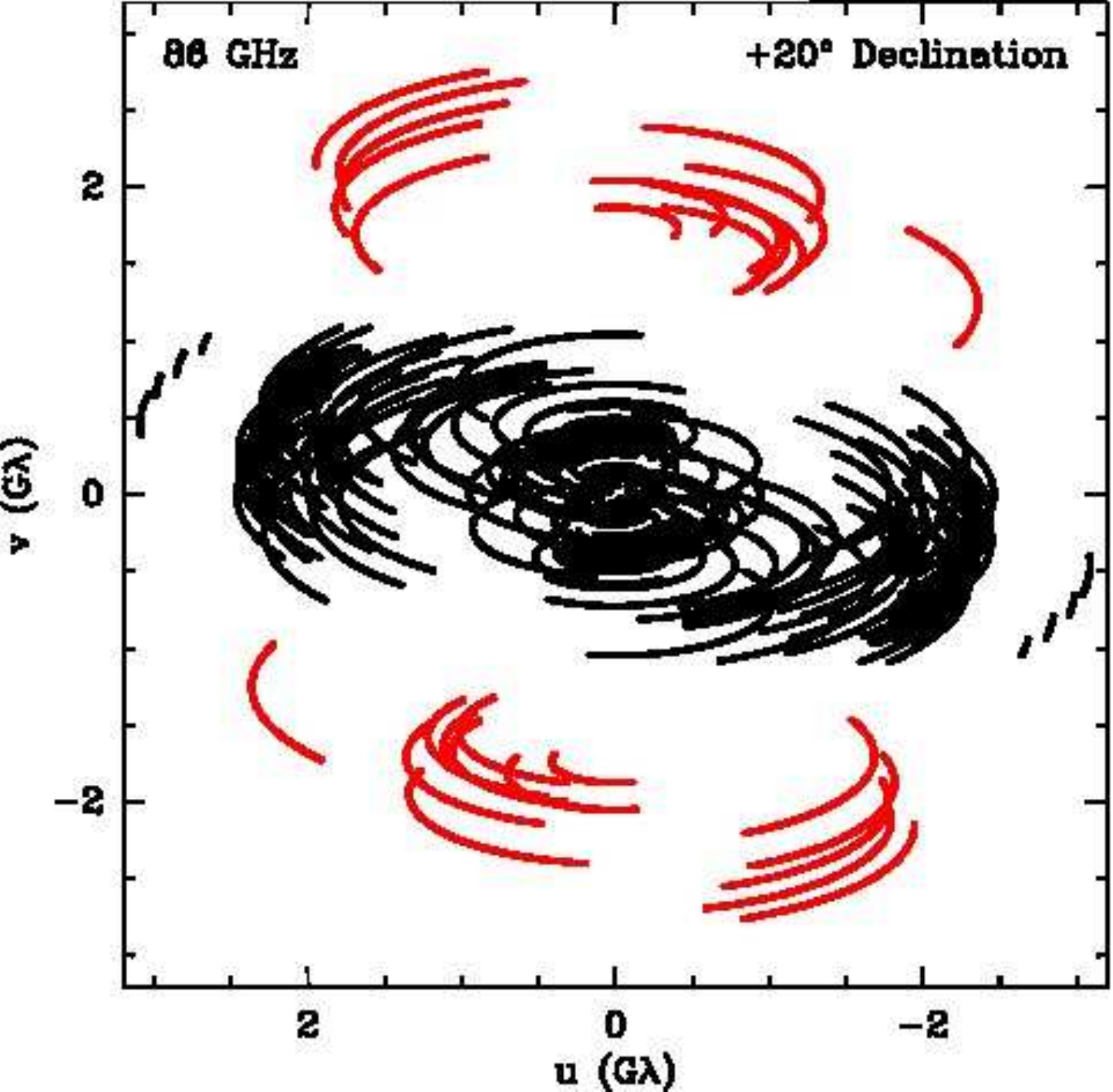}}
\resizebox{0.49\hsize}{!}{\includegraphics{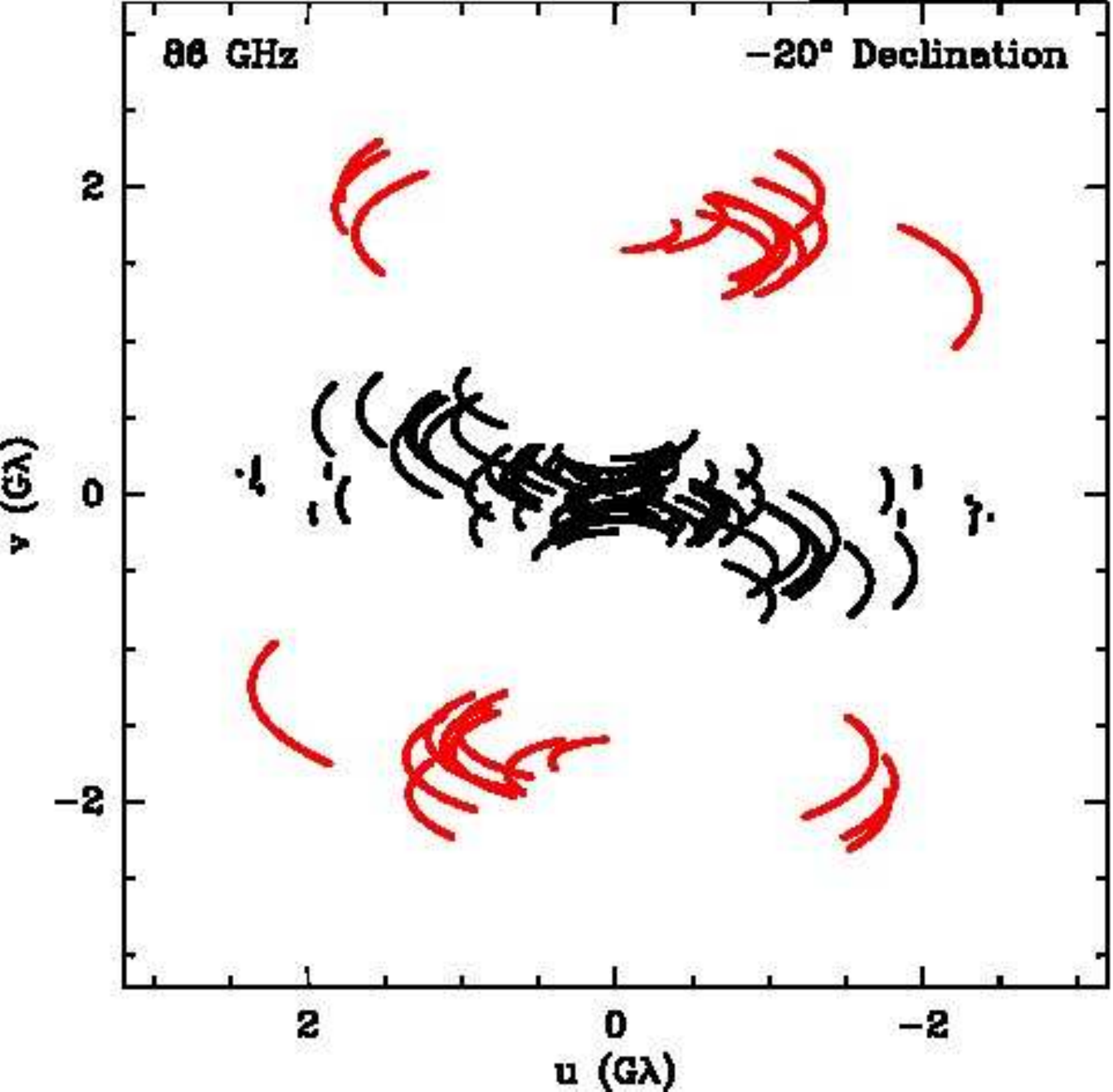}}
\caption{The $(u,v)$ coverage of sources at $+20\degr$ (left) and
  $-20\degr$ (right) declination.  The top row shows the global VLBI
  baseline coverage at 7~mm, and the bottom row shows the coverage at
  3~mm.  Baselines to ALMA are indicated in red.  ALMA provides
  substantial improvement in the north-south coverage of global VLBI
  arrays, especially at 3~mm and for southern sources.
\label{fig-uvcoverage}}
\end{figure}

ALMA will form very long baselines to existing VLBI stations,
improving angular resolution.  ALMA will more than double the
north-south resolution of present 3 and 7~mm VLBI arrays
(Figure~\ref{fig-uvcoverage}).  At higher frequencies, the fringe
spacings of the longest baselines to ALMA are smaller than 20~$\mu$as.
Importantly, ALMA will provide not only increased angular resolution
but also $(u,v)$ coverage that is critical for imaging at 230~GHz and
above.

\section{Black Holes on Event Horizon Scales}\label{blackholes}

As the most extreme self-gravitating objects predicted by Einstein's
General Theory of Relativity (GR), black holes lie squarely at the
intersection of astronomy and physics.  They are now believed to
reside at the heart of the vast majority of galaxies, and power a
subset of X-ray binaries.  Black holes play a critical role at the
endpoints of stellar evolution, as the engines of gamma-ray bursts
(the brightest events in the universe), in the formation and evolution
of galaxies via feedback, and in powering the high-energy phenomena
associated with active galactic nuclei (e.g., launching relativistic
jets and producing the extraordinary electromagnetic luminosities
observed.)

Resolving the immediate environment of a black hole on scales of the
event horizon promises to provide sorely needed empirical insight into
the impact of the black hole upon all of these processes.  It is in
this near-horizon environment where the accretion luminosity and
relativistic outflows are produced, with potential implications on
galactic and intergalactic scales.  Realization of this goal requires
angular resolution that only (sub)millimeter VLBI can provide (Falcke
et al.\ 2000).  The black hole with the largest apparent angular size,
Sgr~A*, subtends tens of \emph{micro}arcseconds.  However, successful
observations using the technique of 1.3~mm VLBI have made it clear
that imaging the emission at the event horizon is now within reach.

The most compelling evidence for this exciting possibility is the
detection of structure on scales of $4~r_\mathrm{Sch}$ (Schwarzschild
radius) in Sgr~A*, the $\sim 4$~million $M_\odot$ black hole at the
center of our galaxy (Doeleman et al.\ 2008).  These results were
obtained using a 3-station 1.3~mm VLBI array, and continued
observations with an enhanced 1.3~mm VLBI network have now measured
time variability on $r_\mathrm{Sch}$ scales in Sgr~A* (Fish et
al.\ 2011) and revealed structure at the very base of the relativistic
jet in M87 (Doeleman et al.\ 2012).  These results confirm the power
of (sub)millimeter VLBI to resolve regions at the black hole boundary
where strong gravity is dominant, creating an opportunity of immense
scientific potential.

Existing 1.3~mm and 0.8~mm telescopes have in recent years engaged in
an effort to produce a network of telescopes to be used for millimeter
and submillimeter VLBI.  Transforming this array into a
wide-bandwidth, short-wavelength VLBI network that spans the globe,
will achieve the resolution, sensitivity, and baseline coverage
required to make true images of the event horizon.  A critical step
will be the inclusion of ALMA as a phased and beamformed single
element in this observing array, the technical project described in
this document.  Parallel efforts to wide-band VLBI instrumentation to
match ALMA bandwidths are funded and underway at other potential
network sites.

\begin{figure*}
\begin{center}
\resizebox{0.3\hsize}{!}{\includegraphics{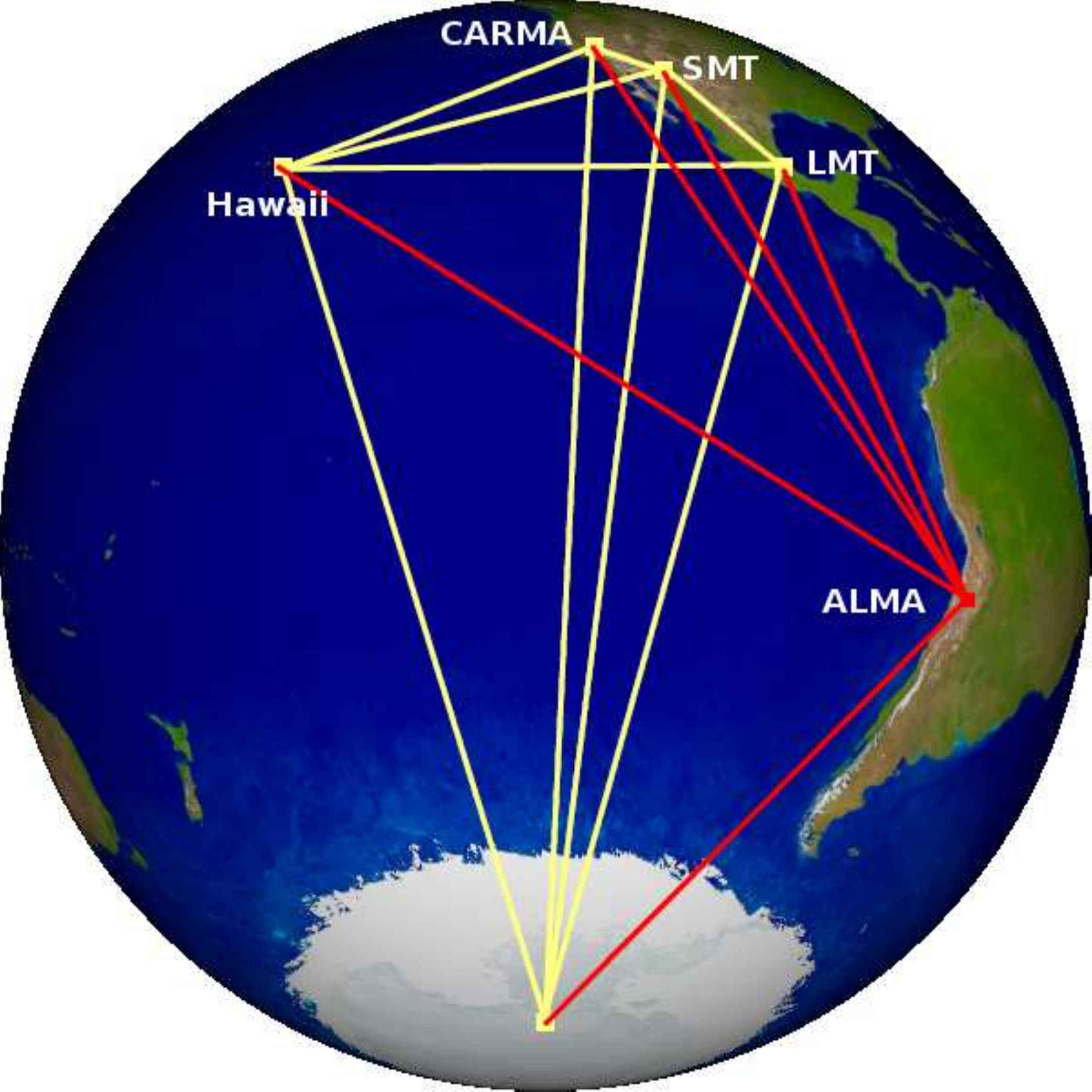}}
\resizebox{0.3\hsize}{!}{\includegraphics{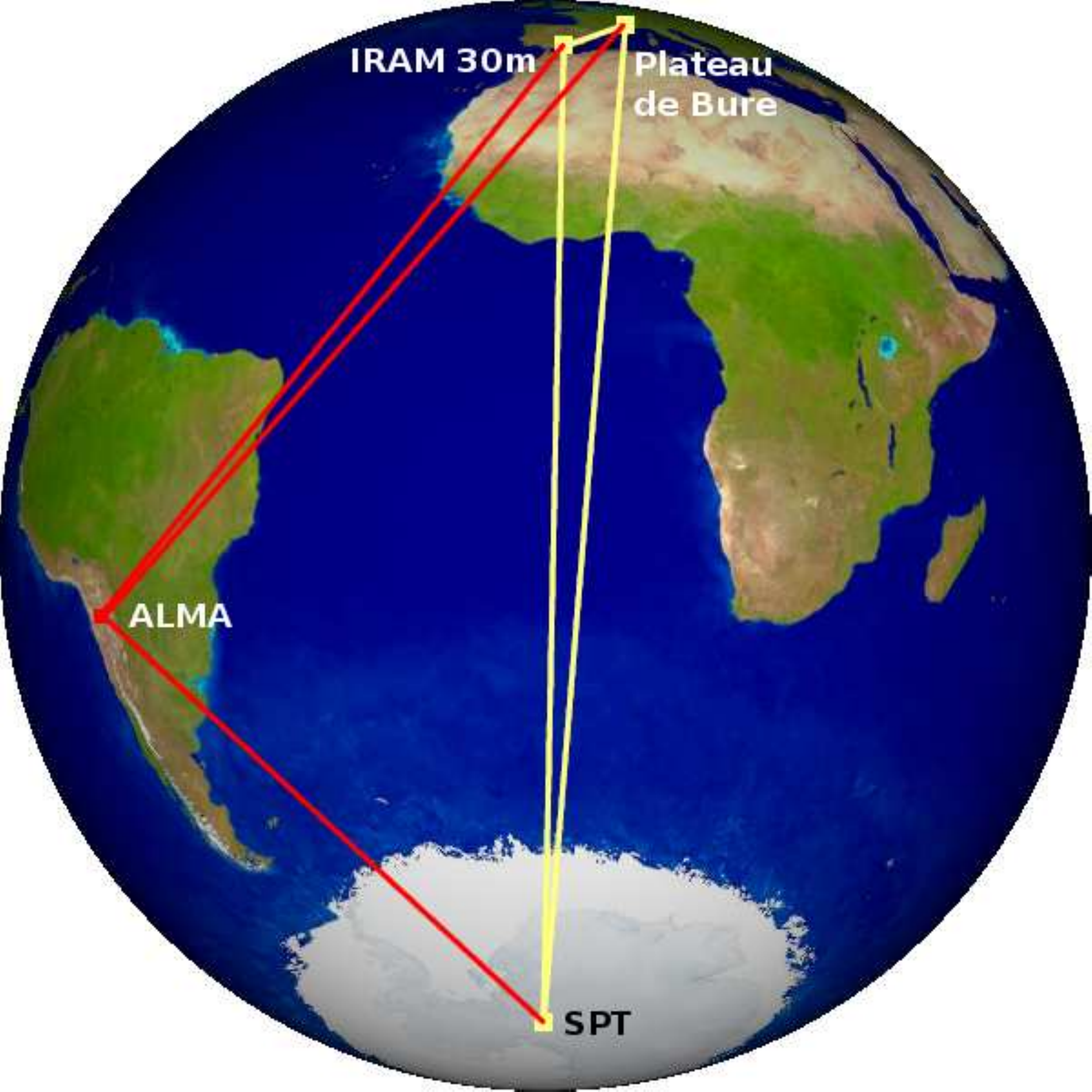}}
\resizebox{0.3\hsize}{!}{\includegraphics{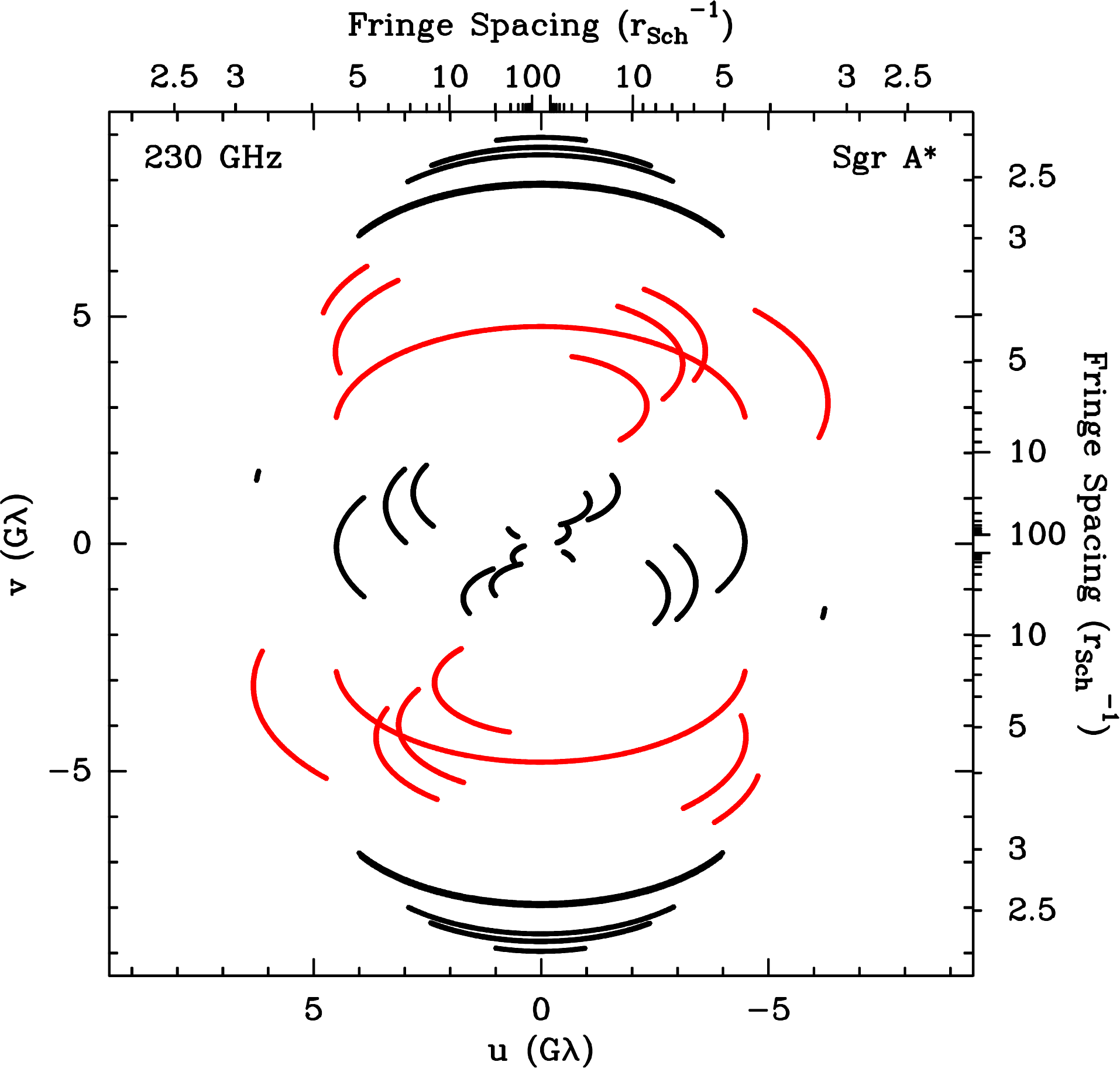}}
\end{center}
\caption{Left and middle: The potential 1.3~mm VLBI network as viewed
  from the declination of Sgr~A*.  Right: The corresponding $(u,v)$
  coverage, with fringe spacings marked in units of $r_\mathrm{Sch}$.
  Baselines to ALMA are marked in red in all panels.
\label{fig-globes}}
\end{figure*}

This combination of phasing ALMA and increased bandwidth will advance
three important areas that will dramatically enhance the current
1.3~mm VLBI array and enable future 0.8~mm VLBI.
\begin{description}
\addtolength{\itemsep}{-1ex}
\item[Sensitivity] Recording up to 8~GHz of bandwidth and coherently
  phasing ALMA will lead to more than an order of magnitude
  sensitivity increase over current 1.3~mm VLBI arrays.  One of the
  greatest impacts of this extraordinary leap in sensitivity will be
  the reliable measurement of interferometric phase, leading to first
  imaging capability.

\item[Polarization] Focused efforts to enable dual-polarization
  recording at 1.3~mm and 0.8~mm VLBI telescopes will allow
  full-polarization VLBI data sets (all four Stokes parameters).  This
  capability will lead to new areas of research that target magnetic
  field structure near the event horizon.

\item[Baseline Coverage] Addition of new 1.3~mm and 0.8~mm VLBI sites
  over the next three years will increase imaging fidelity,
  sensitivity, and temporal coverage of both Sgr~A* and M87.  The
  scheduled addition of ALMA (2015), and inclusion of the South Pole
  Telescope (now a fully-funded project) and the proposed Greenland
  Telescope will provide important new high resolution baselines that
  enable imaging and sensitive tests for time variable
  $r_\mathrm{Sch}$-scale structures (Figure~\ref{fig-globes}).
\end{description}

These new capabilities have the potential to fundamentally transform
our understanding of black holes on event horizon scales.  Due to
their mass and proximity, the two prime targets for such
high-resolution studies are the black holes in the centers of the
Milky Way (Sagittarius A*) and Virgo A (M87).  At a distance hundreds
of times smaller than that to the next nearest supermassive black hole
(SMBH), Sgr~A* can be studied in unparalleled detail and therefore
plays an important role in astrophysics.  On scales much larger than
those probed by VLBI, the world's most powerful optical/IR telescopes
have been trained on Sgr~A* for years to directly observe the orbits
of stars around the black hole and thereby measure its mass and
distance (Ghez et al.\ 2008; Gillessen et al.\ 2009a).  From these and
other observations, the case for linking Sgr~A* with a $\sim 4 \times
10^6~M_\odot$ SMBH is extremely strong (Reid 2009a and references
therein).  Among the most compelling pieces of evidence is the new
intrinsic size of Sgr~A* measured using 1.3~mm VLBI, which implies a
central mass density in excess of $9.3 \times 10^{22}
M_\odot$\,pc$^{-3}$, ruling out all but the most exotic black hole
alternatives (Maoz 1998).  At a distance of 8~kpc, the Schwarzschild
radius of this black hole subtends $r_\mathrm{Sch} \approx 10~\mu$as
(corresponding to an \emph{apparent horizon diameter} of $\sim
50~\mu$as), making the apparent size of its event horizon the largest
that we know of.  VLBI at 1.3~mm wavelength can ``see through'' the
interstellar medium that scatter broadens this source with a
$\lambda^2$ dependence.  Unlike Sgr~A*, the giant elliptical galaxy
M87 exhibits a relativistic jet from sub-parsec to kiloparsec scales
and is possibly the best candidate for the study of jet formation and
collimation on small scales with VLBI (Kovalev et al.\ 2007; Ly et
al.\ 2007; Hada et al.\ 2011, 2013). At a distance of 16.7~Mpc, the
$\sim 6.4 \times 10^9~M_\odot$ central black hole (Gebhardt et
al. 2011) has $r_\mathrm{Sch} \approx 7.5~\mu$as, only slightly
smaller than that of Sgr~A*.

\begin{figure*}[t]
\resizebox{\hsize}{!}{\includegraphics{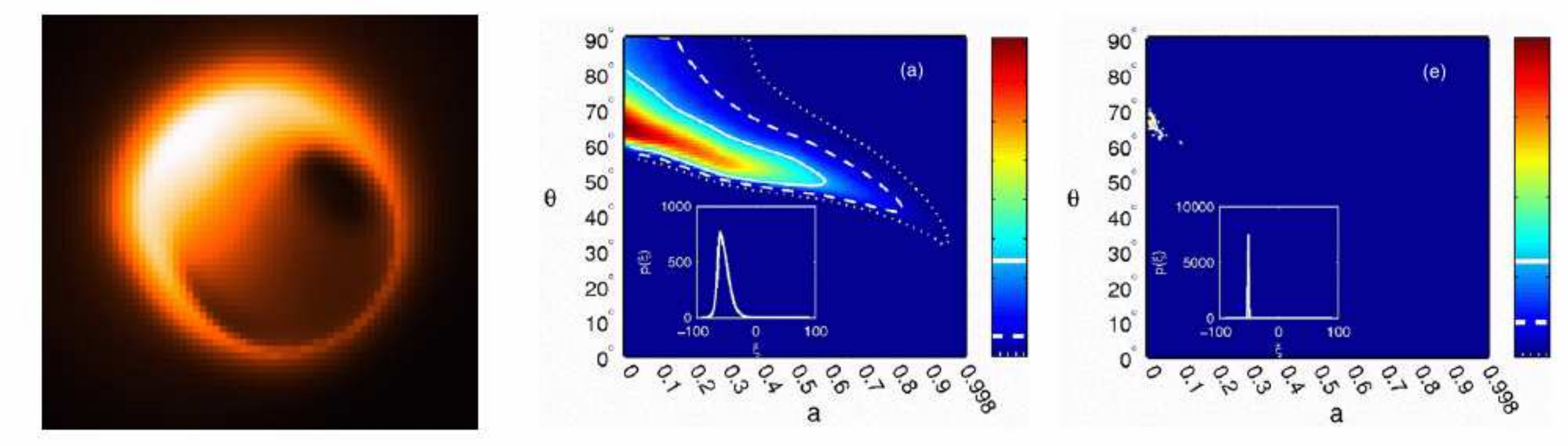}}
\caption{Left: The most likely RIAF model of the accretion flow in
  Sgr~A* (Broderick et al.\ 2011a).  Middle: Probability contours for
  models parameterized by spin magnitude (abscissa) and inclination
  (ordinate) are shown in color, with the marginalized probability for
  the position angle in the plane of the sky indicated in the inset
  (Broderick et al.\ 2011b).  Right: The probability contours tighten
  up dramatically when predicted closure phases from phased ALMA are
  included, and all three components of the black hole spin vector can
  be determined with great precision.
\label{fig-riaf}}
\end{figure*}

\subsection{Constraining the Spin and Viewing Angle of Sgr~A*}

Sgr~A* is highly underluminous, with a bolometric luminosity of around
10$^{-8}$ times the Eddington limit.  The emission from Sgr~A* is
conventionally modelled as arising from a radiatively inefficient
accretion flow (RIAF) in which the electron and ion temperatures are
decoupled from one another (Narayan et al.\ 1995).  The cool electrons
are incapable of radiating much heat, while the hotter ions disappear
through the event horizon or drive outflows (Yuan et al.\ 2003).
Contributions from a jet component are also possible (Falcke et
al.\ 1993).

Some physical constraints on RIAF models arise from fitting the
multiwavelength properties of Sgr~A*, including the spectral energy
density and the implied accretion rate from Faraday rotation and
depolarization at millimeter wavelengths.  However, spatially-resolved
observations are necessary for unambiguously determining properties
such as the spin of the black hole and the viewing geometry---indeed,
for validating the RIAF model in general.  VLBI data at 1.3~mm
(Doeleman et al.\ 2008; Fish et al.\ 2011), which provide this
much-needed resolution, have successfully been used to constrain these
parameters, establishing that the spin vector of the black hole has a
large inclination to the line of sight and identifying a clearly
preferred orientation in the plane of the sky (Broderick et al.\ 2009,
2011a).  More sophisticated general relativistic magnetohydrodynamic
models of the emission (Mo\'{s}cibrodzka et al.\ 2009; Dexter et
al.\ 2010) and even jet models (Markoff et al.\ 2007) reach consistent
constraints, suggesting that the observed emission morphology is
dominated by the geometry of the flow and the general relativistic
beaming and lensing effects near the black hole rather than by
turbulent microphysics in the accretion flow.

While the present three-station VLBI array has been very successful in
determining the structure of Sgr~A*, some parameter degeneracies
persist, largely because the marginal signal-to-noise ratio (S/N) of
the VLBI data prevents use of VLBI phase information.  The highly
sensitive, very long baselines to phased ALMA will enable precise
measurement of the closure phase, the sum of interferometric phase
around a closed triangle of baselines.  Inclusion of closure phase
decisively removes symmetry-flip degeneracies that persist when using
only VLBI amplitudes (Broderick et al.\ 2011b).  Simulations
(Figure~\ref{fig-riaf}) show that measurement of closure phase on
triangles including ALMA will produce exceptionally tight constraints
on the spin vector of the Sgr~A* black hole within the context of RIAF
models.

\begin{figure*}
\resizebox{0.34\hsize}{!}{\includegraphics{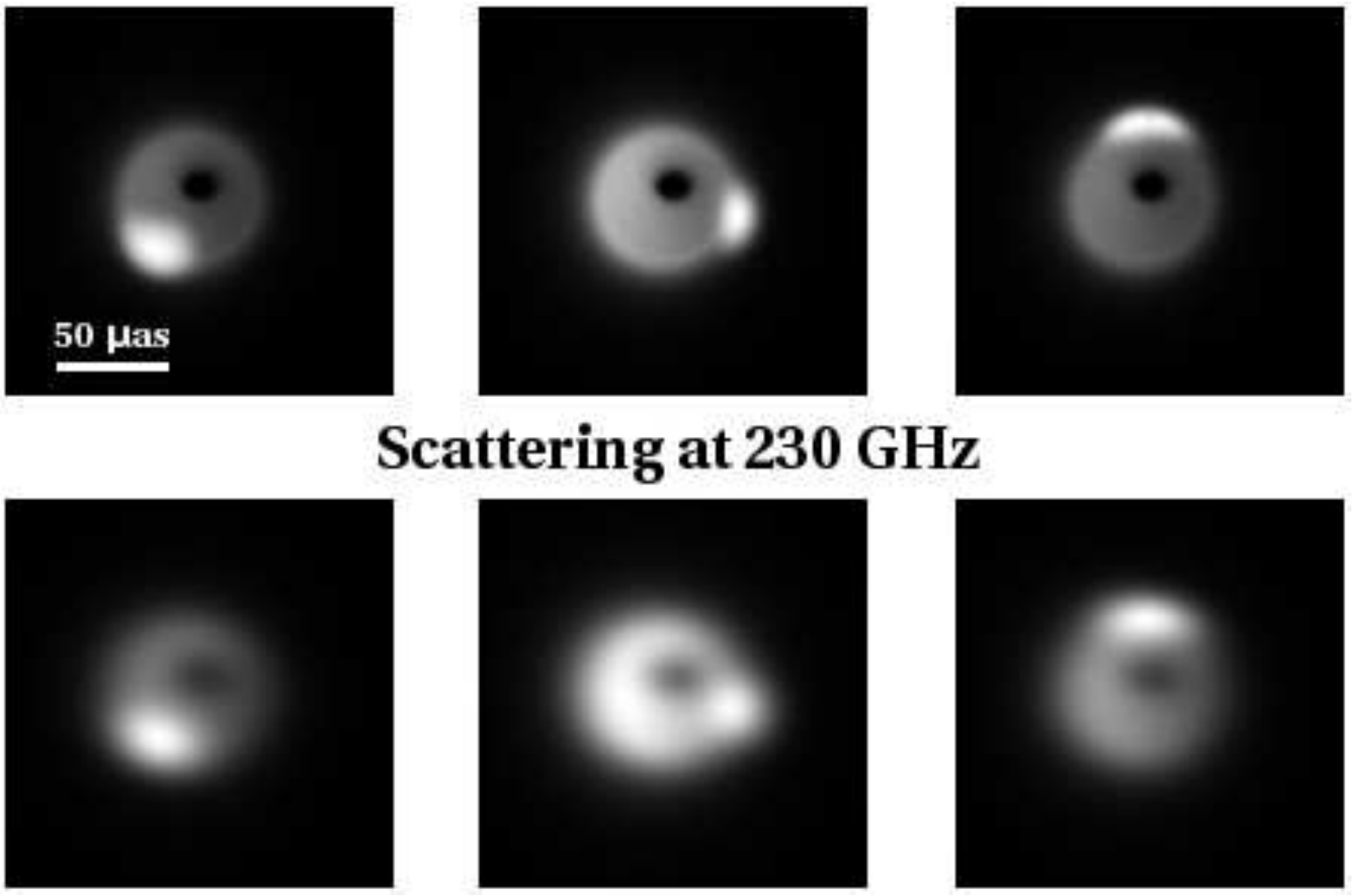}}
\resizebox{0.32\hsize}{!}{\includegraphics{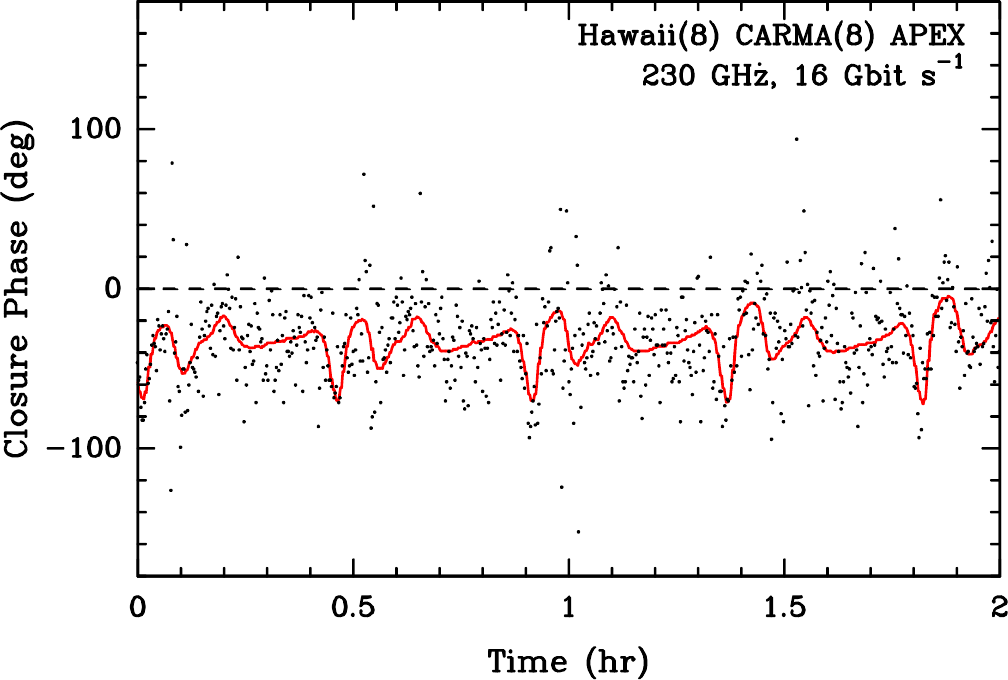}}
\resizebox{0.32\hsize}{!}{\includegraphics{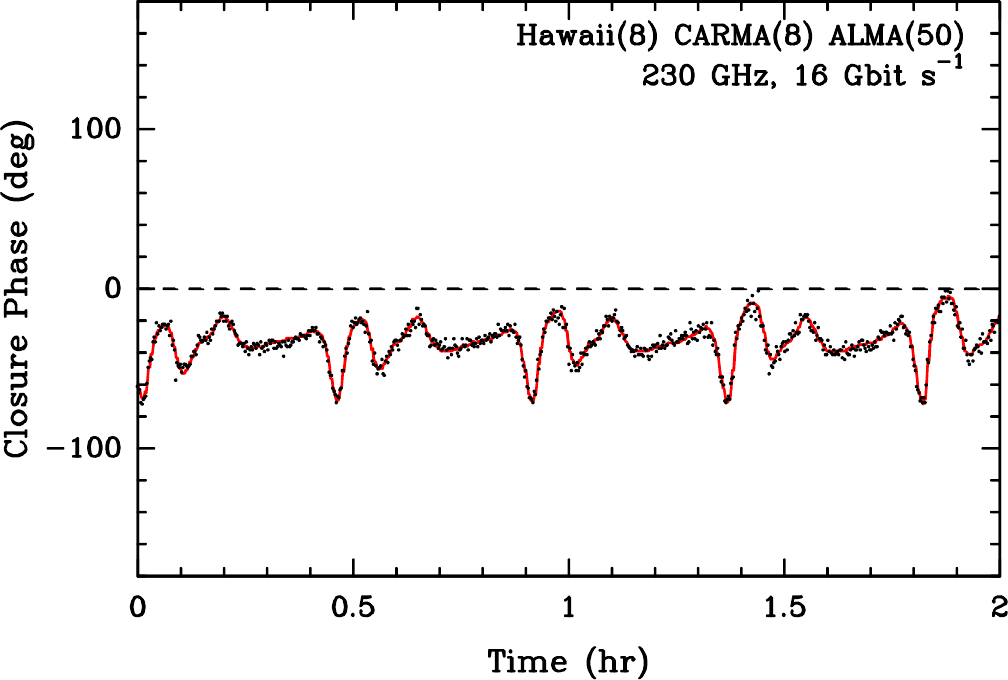}}
\caption{Signature of a hot spot orbiting a black hole at a radius of
  3~$r_\mathrm{Sch}$ with a period of 27 minutes.  The model is shown
  at three equally-spaced orbital phases, with an without interstellar
  scattering.  The plots show the expected closure phases on a
  triangle of stations including the phased SMA, phased CARMA, and
  either a single telescope in Chile (e.g., APEX, middle) or phased
  ALMA (right).  The red curve shows the expected closure phase signal
  in the absence of noise, and the black points show simulated data a
  time cadence of 10~s (comparable to the coherence time of the
  atmosphere).  The substantial sensitivity improvement provided by
  phased ALMA will allow rapid time-domain studies of the variable
  accretion flow in Sgr~A*.  (Figure adapted from Doeleman et
  al.\ 2009)
\label{fig-orbits}}
\end{figure*}

\subsection{Time-Domain Studies}

One of the most promising areas where VLBI can make new contributions
to the study of black hole physics is in searching for time-variable
structures due to inhomogeneities in the accretion flow surrounding
Sgr~A*.  Localized heating in the inner accretion flow is a natural
consequence of magnetic turbulence (Broderick \& Loeb 2006; Dexter et
al.\ 2010) and can give rise to orbiting ``hot spots'', which have
been used to explain the pronounced X-ray, near infrared and
submillimeter flares in Sgr~A* (Yusef-Zadeh et al.\ 2006; Eckart et
al.\ 2006; Marrone et al.\ 2008).  VLBI cannot image these
time-variable structures since they would be smeared out over a single
observing epoch---for Sgr~A*, the innermost stable circular orbit
(ISCO) has a period of 30~min for a non-spinning black hole ($a = 0$),
and only 4~min for one that has maximum spin ($a = 1$).  However,
clear signatures of hot spots should be detectable by using closure
phase.  (For the more massive black hole in M87, the ISCO period is
longer than an observing night, with the result that standard
Earth-rotation aperture-synthesis imaging can be used to make movies
during periods when the emission is variable.)

Simulations of such hot-spots that include full GR ray-tracing and
radiative transfer (Doeleman et al.\ 2009; Fish et al.\ 2009b) show
that baselines to ALMA can not only detect such hot spots using
closure phase, but will have sufficient sensitivity to detect
periodicity if the hot spots persist for multiple orbits in the
accretion flow.  Detection of periodicity would result in a new way to
measure black hole spin and test the validity of the Kerr metric by
studying ensembles of hot-spot orbits over many observing epochs.

Increased bandwidths combined with dual polarization receivers will
also allow high sensitivity VLBI polarimetry of Sgr~A*. At 230 and
345~GHz, the measured low angular resolution linear polarization
fraction of Sgr~A* is $\sim 10$\% (Aitken et al.\ 2000; Bower et
al.\ 2003; Marrone et al.\ 2006, 2007), but models of magnetized
accretion flows predict polarization fractions in excess of 30\% and
up to 60\% on angular scales probed by 230 GHz VLBI (Bromley et
al.\ 2001; Broderick \& Loeb 2006).  It is probable that all previous
polarimetry of Sgr~A* has suffered from significant beam
depolarization.  Only (sub)millimeter VLBI polarimetry, particularly
at the 345~GHz where source opacity is lowest and the blurring of
interstellar scattering is least important, can observe the
small-scale polarized structures that models predict, providing a very
powerful new diagnostic of hot spots and accretion emission.  On
sensitive single VLBI baselines, weaker cross-polarized detections
(e.g., between the right and left circular polarizations, RCP-LCP) can
be phase-referenced to the stronger parallel detections (e.g.,
RCP-RCP), allowing searches for time-variable polarized emission on
$r_\mathrm{Sch}$ scales.

\subsection{Testing General Relativity with (Sub)Millimeter VLBI}

Strong lensing due to general relativity around a black hole is
predicted to produce detectable observational signatures.  When a
black hole is surrounded by an optically thin plasma, lensing should
produce a bright photon ring with a dim ``shadow'' in its interior
(Falcke et al.\ 2000).  The shape of this shadow very closely
approximates a circle for values of the spin $a \lesssim 0.9$
(Takahashi 2004; Johannsen \& Psaltis 2010b).  Confirmation of this
prediction would probe more deeply and cleanly into a relativistic
potential than has ever been done before.

General relativity predicts not only the existence of the black hole
shadow but also its size, which is proportional to the mass of the
black hole.  Some uncertainties remain regarding the mass of the black
hole in Sgr~A* and its distance from the Earth (e.g., Ghez et
al.\ 2008; Gillessen et al.\ 2009a, 2009b), although the predicted
diameter of the shadow is approximately 50~$\mu$as.  A wide variety of
models of the 1.3~mm emission, including models in which this emission
arises in an accretion disk and others in which it originates from a
jet, predict that the morphology of the emission will be dominated by
general relativistic lensing, resulting in crescent-like images and a
shadow that will be detectable on VLBI baselines to ALMA (e.g., Dexter
et al.\ 2010).  Similar results may be found for M87, although there
is less agreement on the predicted morphology of the emitting region
(e.g., Dexter et al.\ 2012).  Conversely, if general relativity is
assumed to be true, measuring the size of a black hole shadow provides
an independent measurement of the ratio of the black hole mass to the
source distance (Johannsen et al.\ 2012; Ruprecht 2012).

\begin{figure}
\resizebox{\hsize}{!}{\includegraphics{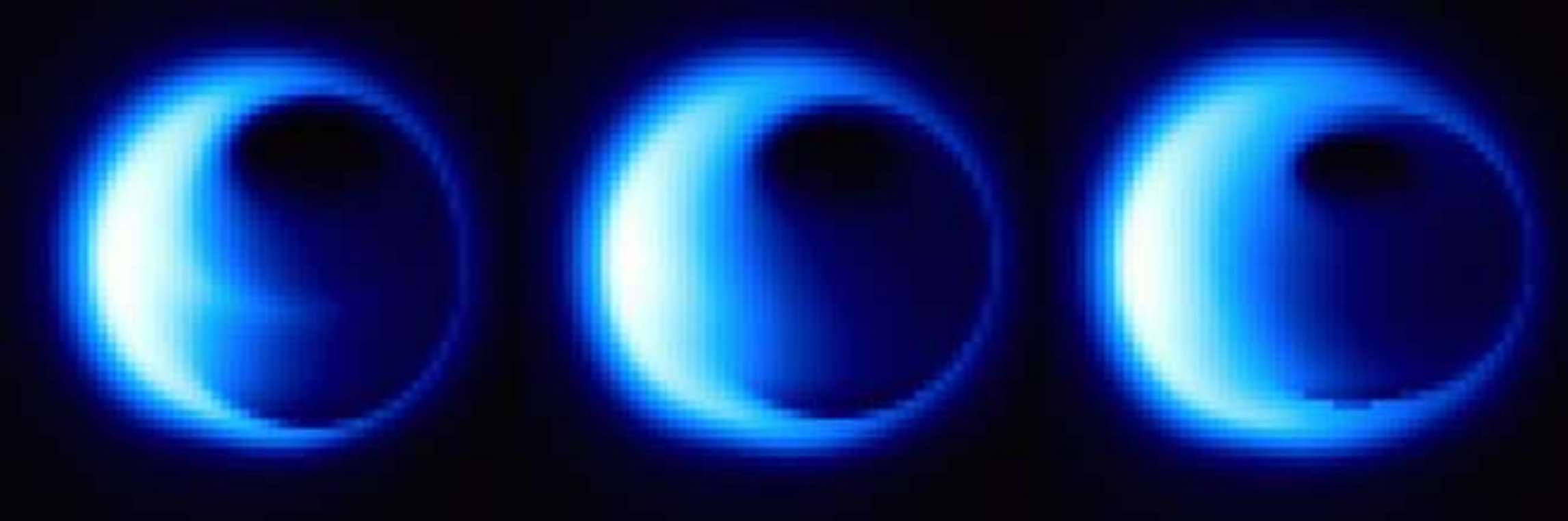}}\\
\resizebox{\hsize}{!}{\includegraphics{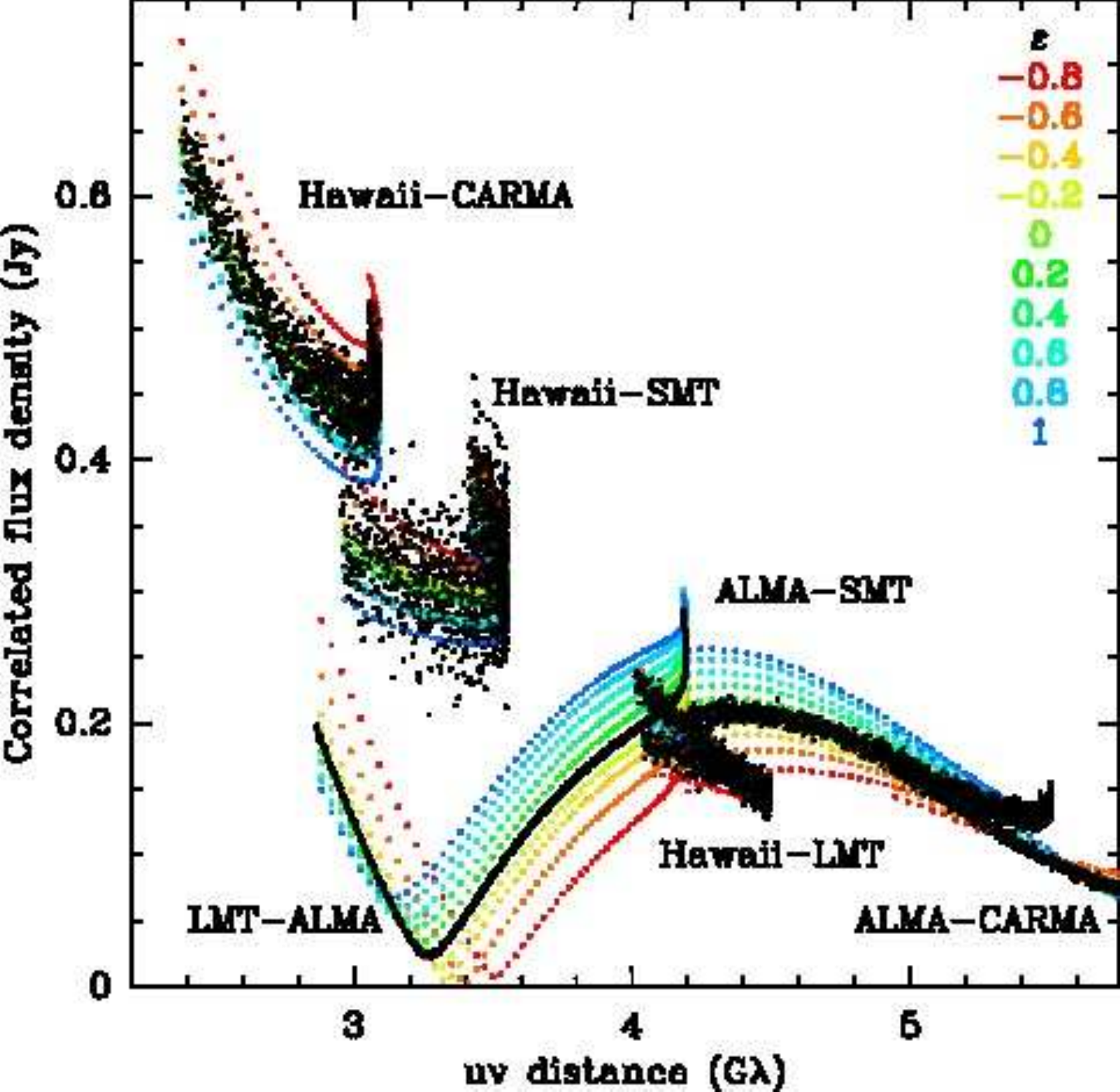}}
\caption{In spacetimes that do not obey the no-hair theorem near a
  black hole, the shadow can be noncircular.  From left to right, the
  three top panels show model images (Broderick, A.E., Johannsen, T.,
  Loeb, A., \& Psaltis, D., in preparation) with $\epsilon = -0.8$,
  $\epsilon = 0$ (general relativity), and $\epsilon = +1$.  These
  models predict measurably different amounts of correlated flux
  density on baselines between ALMA and North America (bottom, with
  simulated data overplotted).
\label{fig-epsilon}}
\end{figure}

The shape of the photon ring also provides a test of the ``no-hair''
theorem, which states that the exterior spacetime of a black hole can
be completely characterized by the black hole's mass, spin, and
electric charge.  Since it is difficult to envision how a black hole
could sustain a gravitationally relevant net electric charge (see
Blandford \& Znajek 1977), the only independent multipole moments of
the spacetime of a real astrophysical black hole are the monopole and
dipole moments, which correspond to the black hole mass and spin,
respectively.  In general relativity, all higher moments are
expressible in terms of the mass and spin alone.  The simplest
theoretical spacetimes that violate this theorem assume a quadrupole
moment $Q = -M \left(a^2 + \epsilon M^2\right)$, where $a$ is the
black hole spin, $M$ is the mass, and $\epsilon$ parameterizes the
deviation from general relativity (Glampedakis \& Babak 2006;
Johannsen \& Psaltis 2010a).  When a nonzero quadrupole deviation is
introduced, the shape of the photon ring and shadow becomes
noncircular (Johannsen \& Psaltis 2010b), imprinting a signature onto
VLBI quantities that would be detectable with an observing array that
includes phased ALMA (Figure~\ref{fig-epsilon}).  Similarly, many
other alternative theories of gravitation make testable predictions
for the properties of black hole shadows (e.g., the review by Falcke
\& Markoff 2013).

\subsection{Observing Jet Launching on Schwarzschild Radius Scales}

Jets are believed to be powered through extraction of energy from
accreting supermassive black holes.  Feedback from these jets enriches
and heats the intergalactic medium and plays a role in galaxy
formation.  Mechanisms have been proposed to explain how energy and
angular momentum can be extracted from the accretion flow or the spin
of the black hole (e.g., Blandford \& Znajek 1977; Blandford \& Payne
1982), and general relativistic magnetohydrodynamic simulations are
beginning to be able to model the magnetized plasmas in the inner
accretion and outflow region (e.g., McKinney et al.\ 2012), but the
lack of observations probing these small angular scales make it
difficult to distinguish among classes of models that have the same
behavior on large scales.

\begin{figure}[t]
\resizebox{\hsize}{!}{\includegraphics{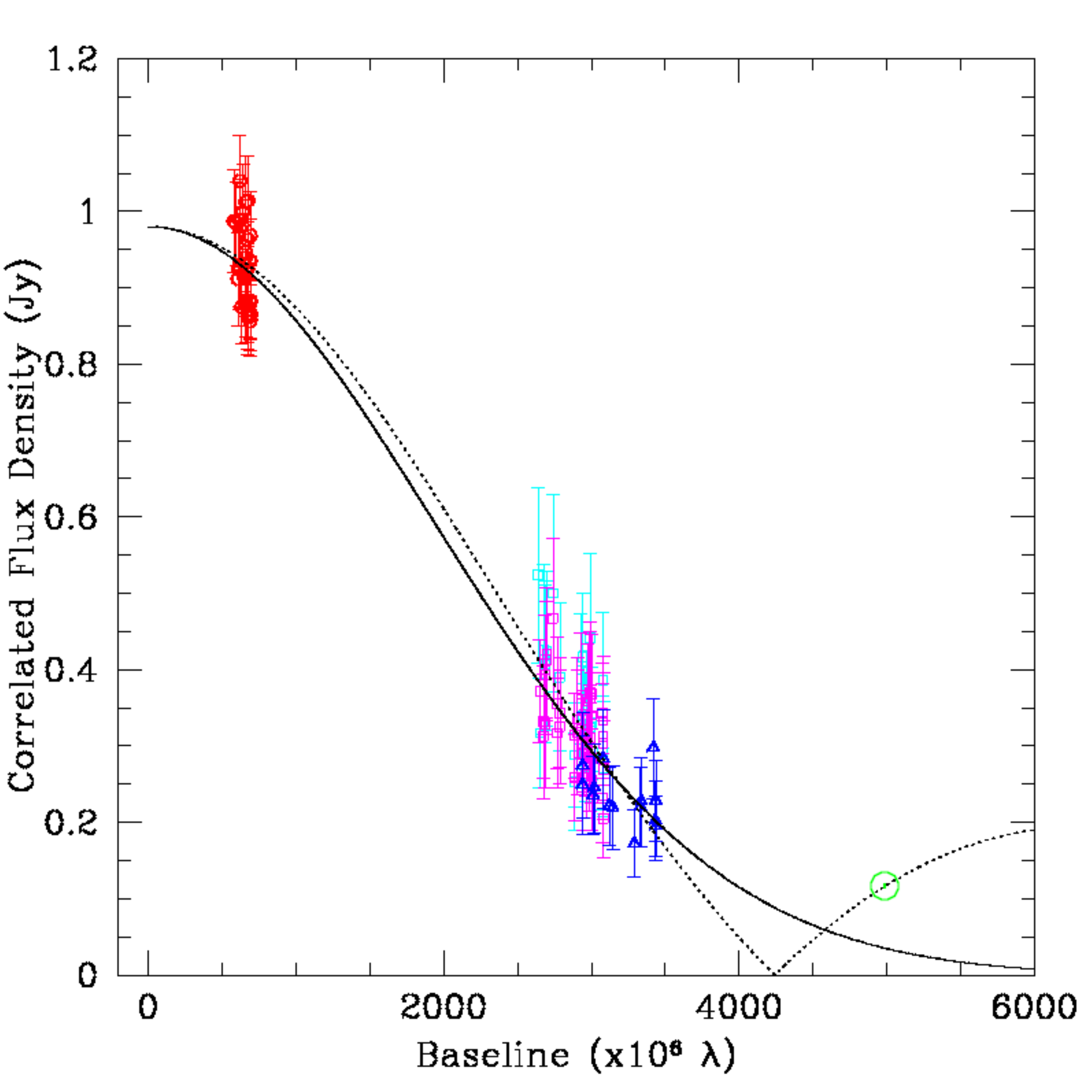}}
\caption{Detections of M87 with 1.3~mm VLBI observations in 2009
  (Doeleman et al.\ 2012).  The solid line shows a Gaussian fit to the
  data, approximating the emission from a forward jet, and the dotted
  line indicates a ring fit, approximating the expected emission from
  a lensed counterjet around the black hole shadow.  The circled green
  point shows the $1\,sigma$ sensitivity on a simulated 10-s data
  point between phased ALMA and the SMT with an effective bandwidth of
  4~GHz in two polarizations.
\label{fig-m87data}}
\end{figure}

\begin{figure}[t]
\resizebox{\hsize}{!}{\includegraphics{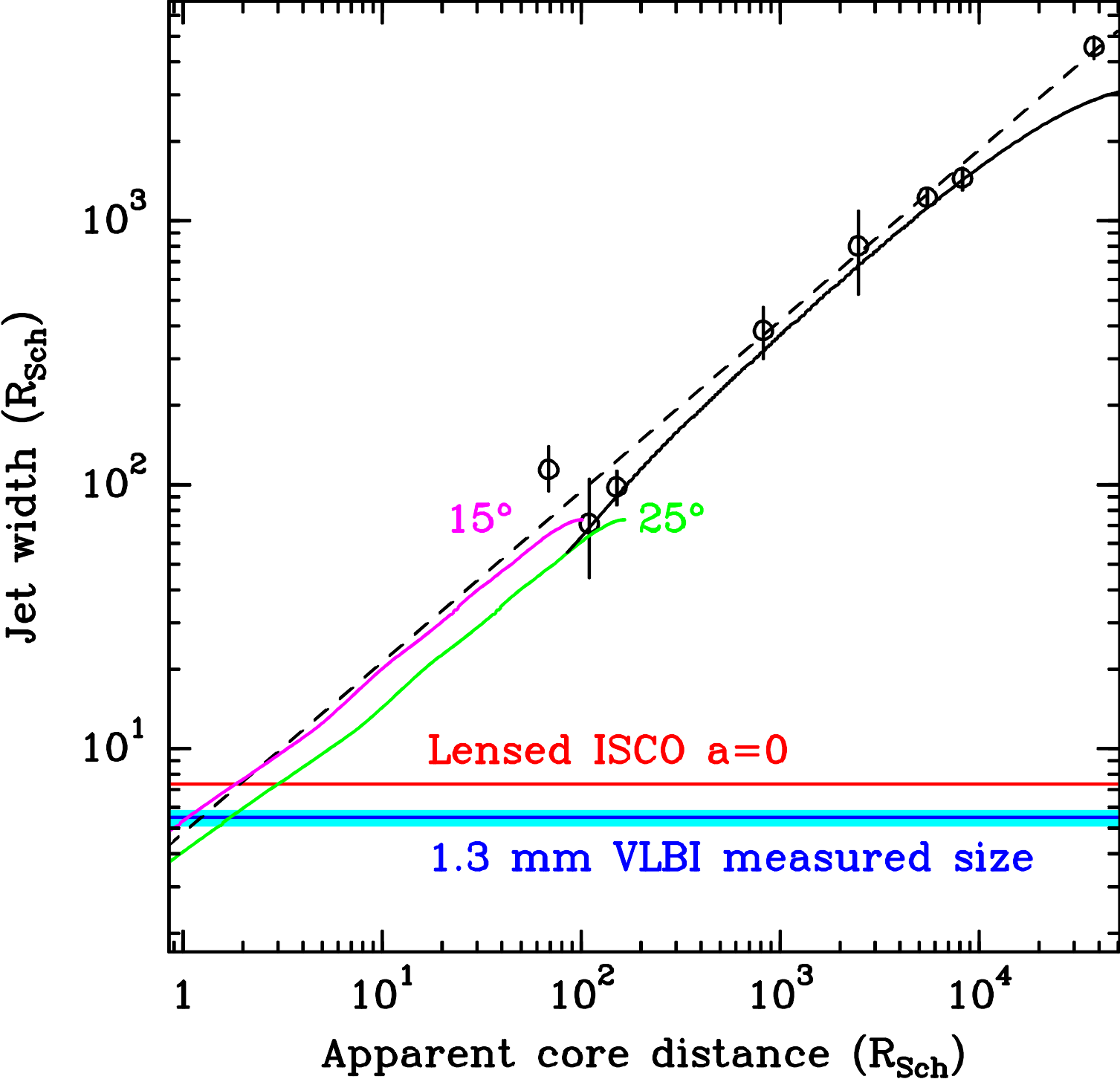}}
\caption{The measured size of the 230 GHz emission in M87 (blue, with
  the cyan region denoting the $3\,\sigma$ error bars) is smaller than
  the lensed size of the innermost stable circular orbit for a
  nonrotating black hole (red).  The size of the emission, which is
  known to originate from within a few Schwarzschild radii of the
  black hole (Hada et al.\ 2011), is consistent with both the observed
  $r^{2/3}$ jet width profile seen at longer wavelengths (symbols,
  from Biretta et al.\ 2002) and the theoretical jet width profile
  (also $r^{2/3}$, shown in magenta and green for inclination angles
  of 15 and 25 degrees; courtesy J. McKinney).  The figure is adapted
  from Doeleman et al.\ (2012).
\label{fig-m87jetwidth}
}
\end{figure}

Recent observations have detected M87 on baselines between Hawaii and
the western US, establishing that at 1.3~mm the inner jet is
approximately 38~$\mu$as (4.8~$r_\mathrm{Sch}$) in size
(Figure~\ref{fig-m87data} and Figure~\ref{fig-m87jetwidth}; Doeleman
et al.\ 2012).  This size favors models in which the jet is launched
within a few Schwarzschild radii of the black hole (e.g., Nishikawa et
al.\ 2005) rather than arising from magnetic fields anchored out past
the ISCO of the disk (e.g., De Villiers \& Hawley 2003).  This size is
also substantially smaller than the ISCO of a nonrotating black hole,
which may suggest that the accretion flow rotates in a prograde sense
around a spinning black hole.  However, further observations on
baselines oriented perpendicular to the roughly east-west direction of
the Doeleman et al.\ (2012) detections, such as those between the US
and ALMA, will be necessary to break degeneracies between this
interpretation and other models.

At longer wavelengths, the inclusion of phased ALMA in VLBI imaging
arrays can address questions relating to jet launch and collimation
mechanisms.  The jet in M87 has been observed to have a parabolic jet
width profile on scales ranging from $10^2$ to $3 \times
10^5~r_\mathrm{Sch}$ (Junor et al.\ 1999; Asada \& Nakamura 2012).
High-fidelity imaging at 3~mm is required to make the connection
between the scales accessible to 1.3~mm VLBI (a few Schwarzschild
radii) and the $100~r_\mathrm{Sch}$ scales observable at longer
wavelengths.  The structure of the M87 jet at 3~mm is observed to be
highly variable, but Global mm-VLBI Array (GMVA) observations suffer
from limitations in image fidelity and sensitivity.  Higher-resolution
jet images would allow a more precise measurement of the jet opening
angle, which is predicted to be at least 60\degr\ if centrifugal
acceleration along poloidal magnetic field lines connected to the
accretion disk is responsible for jet formation (Blandford \& Payne
1982).  Polarimetric observations of the emission near the jet base
would help clarify the three-dimensional geometry of the magnetic
field that launches the jet.  Multi-epoch observations at high
fidelity will allow structures within the jet to be tracked, providing
measurements of the velocity along the jet length.  Motions in the jet
are known to be superluminal at the location of the knot HST-1,
approximately 100 pc ($> 10^5~r_\mathrm{Sch}$) from the black hole
(Biretta et al.\ 1999; Cheung et al.\ 2007), but there is substantial
observational disagreement as to whether the velocity at the jet base
is highly subrelativistic, relativistic, or even superluminal (Kovalev
et al.\ 2007; Ly et al.\ 2007; Acciari et al.\ 2009).  The improved
imaging capability provided by the inclusion of phased ALMA in VLBI
observing arrays will clarify the acceleration profile of the jet in
M87.

\subsection{Imaging Black Holes}

Substantial scientific progress has been made to date understanding
Sgr~A* and M87 with non-imaging techniques, but much of this work has
necessarily been dependent on models of the assumed emission
morphology based on physical intuition and unresolved multiwavelength
observations.  Direct imaging would provide a model-independent
picture of the inner emitting region around supermassive black holes.

The prospects for imaging Sgr~A* or M87 in the next few years are
great.  Bispectral maximum entropy imaging methods originally
developed for optical interferometry (e.g., BSMEM; Buscher 1994) have
shown great promise in being able to produce image reconstructions
from sparsely-sampled $(u,v)$ coverage (e.g., Malbet et al.\ 2010).
Monte Carlo imaging techniques, such as those used by the MACIM code
(Ireland et al.\ 2006), also appear to be particularly well suited to
millimeter VLBI (Ruprecht 2012).  Compared to deconvolution-based
methods, these forward-imaging algorithms are able to provide
significantly better image quality and, in the case of sparse arrays,
often succeed when CLEAN fails to converge at all.

\begin{figure}[t]
\resizebox{\hsize}{!}{\includegraphics{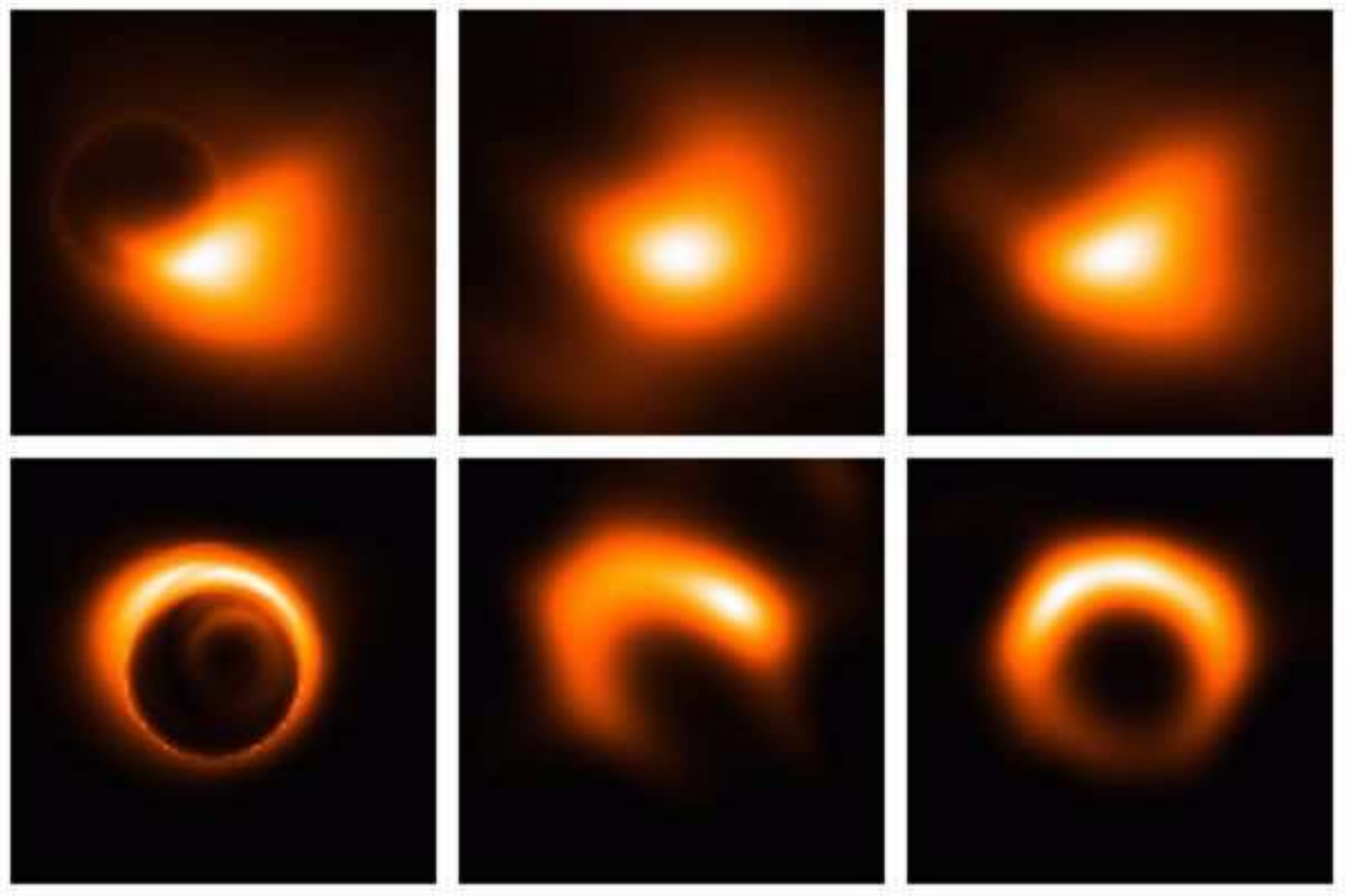}}
\caption{Models and simulated reconstructions of the innermost
  emitting region in M87.  The top left image shows a model of a
  magnetically-dominated jet anchored in an accretion disk around the
  black hole with a footprint $\sim 10~GM/c^2$ (Broderick \& Loeb
  2009).  The bottom left image shows a model in which most of the jet
  emission arises within the photon sphere at 2 to 4~$GM/c^2$, causing
  lensed counterjet emission to dominate in the observer's frame
  (Dexter et al.\ 2012).  Middle panels show BSMEM reconstructions of
  these models from simulated observations using an early array
  consisting only of phased ALMA, phased SMA, phased CARMA, and the
  SMT.  Right panels show reconstructions from this array plus the
  IRAM 30-m telescope, the phased Plateau de Bure Interferometer, and
  the LMT.  The early array can easily distinguish between the two
  classes of models.  The full array can produce reasonable images of
  the emission, possibly providing a clear view of the shadow, from
  which the black hole mass can be measured directly.
\label{fig-m87images}}
\end{figure}

An array consisting solely of phased ALMA and the three telescope
sites in the US that have been participated in 1.3~mm VLBI
observations (CARMA, SMT, Hawaii) can produce images that are
sufficient to distinguish between models of the M87 jet that differ
only on scales of a few $r_\mathrm{Sch}$.  Figure~\ref{fig-m87images}
shows two models, one of which has the jet forming close to the black
hole with its emission strongly lensed into a ``shadow'' morphology.
In the other model, the jet forms farther out from the black hole and
the image is dominated by the approaching outflow, though a faint
shadow is also observed.  The addition of other antennas that are
likely to be VLBI-capable at this wavelength before or shortly after
the completion of the ALMA phasing project will enable much higher
fidelity images to be produced.  This is a possibility that can only
be pursued with the sensitivity and baseline coverage provided by
phased ALMA.

\section{Active Galactic Nuclei and Jet Physics}\label{agn}

Powered by accretion onto black holes with masses up to $10^{10}$
M$_\odot$, AGNs represent the most energetic long-lived phenomenon in
the universe.  In $\sim 10$\% of these objects, a substantial fraction (as
much as tens of percent) of the energy of accretion is converted into
kinetic energy of outflows in the form of relativistic, highly
collimated jets of energetic plasma and magnetic fields.  The jets
emit highly variable nonthermal radiation across the electromagnetic
spectrum that can greatly exceed the thermal emission from the
accretion disk and host galaxy in blazars, AGNs in which one of the
jets points in our direction.  In some cases, the bulk of the observed
flux emerges in $\gamma$-ray photons in the GeV or even TeV energy range.
One of the most intriguing and challenging quests of current
astrophysics is to understand the physical conditions and processes
that give rise to the formation of these relativistic jets, production
of high-energy particles, and emission of $\gamma$-rays.  Of particular
interest is the question of how accretion onto SMBHs generates such
high-powered directed outflows.

AGN jets have profound effects on the surrounding environment, e.g.,
in clusters of galaxies (McNamara et al.\ 2005).  Energy and momentum
exchange between jets and the intracluster medium between galaxies can
regulate the formation and growth of galaxies and the SMBHs within
them, as well as limit the rate of mergers and flows onto galaxies
hosting SMBHs (Springel et al.\ 2005; Di Matteo et al.\ 2005).
Moreover, radiation from the jets ionizes gas along a cone surrounding
the channel of flow within the host galaxy.  Shock waves form where
jets ram into the external medium, inducing large bulk motion of gas
and even triggering star formation.  These effects are visible in AGNs
today, but were even more prevalent at earlier epochs of the universe,
when many galaxies were still in the formation phase and a greater
proportion were strongly active.  Understanding galaxy formation and
the birth of the early generations of stars is therefore intimately
tied to our understanding of how black holes grow by accretion and
generate relativistic jets, and how the jets propagate from the center
of a galaxy to its outskirts and beyond.

\subsection{High-Resolution Jet Imaging}

VLBI is the only technique currently capable of viewing directly the
parsec- and subparsec-scale regions of jets in AGNs.  At frequencies
of 43 and 86~GHz, the jet is mostly optically thin.  The VLBA
currently provides angular resolution of $\sim 100~\mu$as in the
east-west direction at wavelengths of 3 and 7~mm, but only
$\sim250~\mu$as in the north-south direction.

The inclusion of phased ALMA as a VLBI station will more than double
the north-south resolution of 3 and 7~mm VLBI arrays
(Figure~\ref{fig-uvcoverage}).  Such high-resolution observations can
image the acceleration and collimation zone of the flow, which
currently can be explored only in an indirect manner through time
variability of flux and polarization.  The jets of even the brightest
blazars are challenging to detect with arrays composed of single-dish
telescopes operating at 3~mm; adding phased ALMA would improve this
dramatically.

\begin{figure}
\resizebox{!}{1.75truein}{\includegraphics{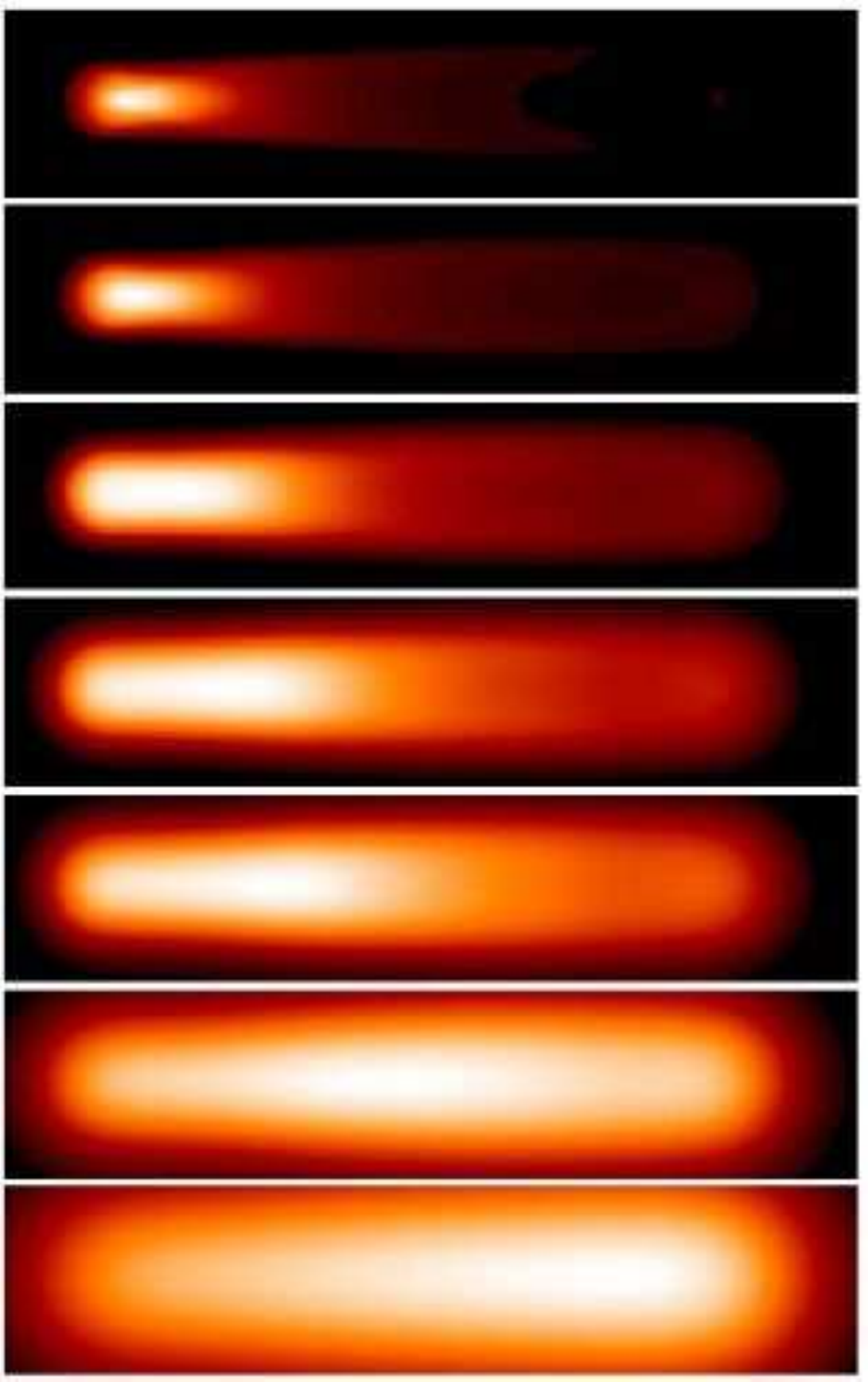}}
\resizebox{!}{1.75truein}{\includegraphics{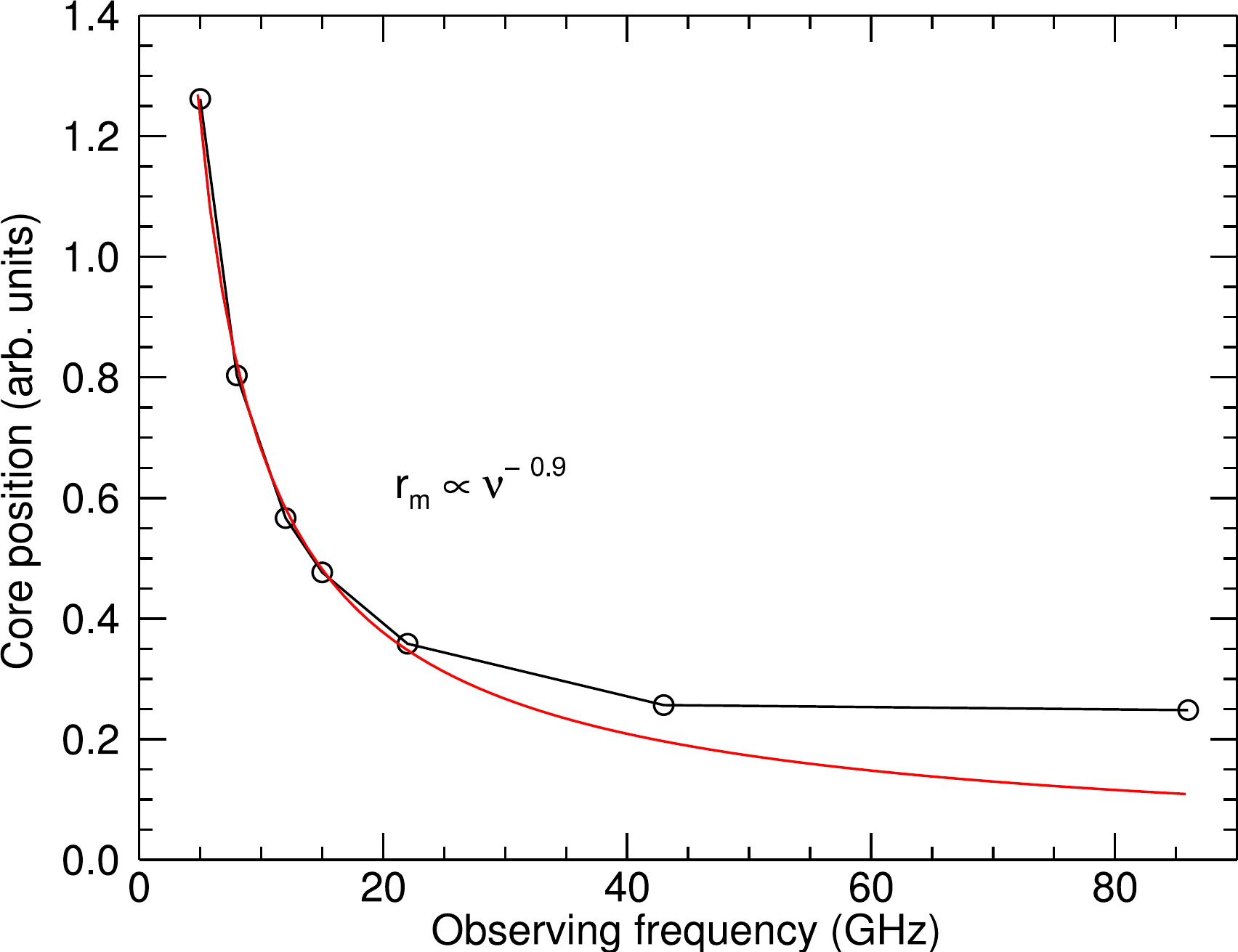}}
\caption{ \emph{Left, top to bottom}: Synchroton models of a blazar
  jet at 86, 43, 22, 15, 12, 8, and 5~GHz (based on G\'{o}mez et
  al.\ 1995, 1997; Mimica et al.\ 2009).  \emph{Right}: The core
  location deviates from a Blandford \& K\"{o}nigl (1979) jet model
  (red curve) at ALMA frequencies, indicative of the location of the
  recollimation shock.
\label{fig:shock}}
\end{figure}

The jets of radio-loud AGNs contain a bright, compact feature at the
upstream end of 7 and 3~mm images.  This is called the core, although
in blazars and blazar-like radio galaxies there is considerable
evidence that it lies more than 0.5~pc from the SMBH that powers the
AGN (e.g., Chatterjee et al.\ 2009, 2011).  The core is a key structure
in a jet.  At centimeter wavelengths what appears to be the core is
generally just the transition between optically thick and thin
emission, but at millimeter wavelengths it has the properties of a
physical structure, such as a standing cone-shaped shock (Cawthorne
2006; D'Arcangelo et al.\ 2007; Marscher et al.\ 2008, 2010).
Synchrotron models show that in some sources the 7 and 3~mm emission
is optically thin enough to trace the recollimation shock
(Figure~\ref{fig:shock}).

A surprising result of recent studies that combine time sequences of
7~mm VLBA images with monitoring of flux across the electromagnetic
spectrum is that, in many blazars, most of the infrared, optical,
X-ray, and $\gamma$-ray emission originates in or near the core.  It
is frustrating that the core is only slightly resolved (and not
resolved at all in the north-south direction) at 7~mm.  Currently, it
is not generally resolved much better at 3~mm because the
trans-continental baselines are too noisy.  With the addition of ALMA,
millimeter VLBI can resolve the cores in bright blazars as well as
image a portion of the jet between the core and the central engine.
By doing so, VLBI studies will address through direct imaging the key
question of how ultra-relativistic jets are generated by accreting
SMBHs.  In particular, such imaging can provide clues regarding how
and where the outflow transforms from purely electromagnetic to
kinetic energy-dominated, and the relative roles played by velocity
shear, turbulence, shocks, and plasma instabilities in the flow.

Some of the most exciting recent discoveries in AGN jets have arisen
from studies of their structure transverse to the flow direction.
Observations of large scale AGN jets with the VLA (Laing \& Bridle
2002) show distinct velocity gradients in which a fast central spine
is surrounded by a slower sheath that interacts with the external
medium.  It is not yet clear to what extent these interactions serve
to stabilize and/or decelerate the flow, and how they influence the
overall evolution and structure of the jet.  Our knowledge is more
limited on small scales, due to the need for high dynamic range
polarimetric imaging, preferably with a symmetric beam.  The addition
of phased ALMA to VLBI arrays will provide significant improvement in
these directions, allowing observations of limb brightening and
changes in polarization that provide powerful probes of the flow
properties.

\subsection{Connections to Optical Imaging}

The European Space Agency's Gaia optical astrometry mission (expected
launch time late 2013) will provide astrometric and photometric
observations of a large sample of QSOs down to V$ = 20$~mag. Its
astrometric accuracy will reach 25~$\mu$as, and the observations of
approximately a half million QSOs will eventually lead to the Gaia
Celestial Reference Frame (GCRF).  It will be possible to establish a
direct connection between the radio-based International Celestial
Reference Frame (ICRF2) and the optical-based GCRF for the first time
(e.g. Charlot \& Bourda 2012).  Gaia is expected to provide
information about the optical emission within 1~mas of the central
engine---within a region of 8~pc of the SMBH at $z \approx 1$ (e.g.,
Browne 2012, and references therein). A recent study (Popovi{\'c} et
al.\ 2012) showed that Gaia will be able to detect variations in the
inner structure of the QSOs via positional offsets of the optical
photocenters.  ``Pre-Gaia'' optical observations of QSOs by Ant\'on et
al.\ (2012) showed photocenter jittering of a few tens of parsecs in
three sources accompanied by optical magnitude variations.  These
findings can be explained by either blobs or shocks traveling in the
jet (very similar what is observed in the radio regime) or by double
variable sources at the center of the objects---e.g., binary
supermassive black holes (Popovi{\'c} et al.\ 2012).  As of now, it is
not clear whether the non-thermal optical and radio emission originate
from the same location.  With the inclusion of phased ALMA in
millimeter VLBI, the radio emitting region can be measured and mapped
even closer to the central engine, and thus the region where the
optical emission measured by Gaia may originate can be mapped with
high fidelity in the radio regime.

Gaia will observe $\sim 500000$ QSOs.  Using a conservative estimate
for the proportion of compact, radio-loud, and millimeter-bright
sources, a couple hundred sources will be observable with the
millimeter VLBI array containing phased ALMA.  This sample of radio
sources thus can be studied by the highest resolution VLBI instrument
in the millimeter regime, to investigate whether the observed optical
characteristics can be connected to features in the jet launching
region revealed in the millimeter regime.

\subsection{VHE Photon Production}

An exciting discovery made by space-borne instruments and ground-based
Cherenkov telescopes is the detection of VHE $\gamma$-rays from
over 1000 AGN jets.  There are currently several competing models for
the source of this radiation, which include upscattering of
synchrotron photons from either the jet (e.g., Bloom \& Marscher
1996), accretion disk (Dermer \& Schlickeiser 1994), broad
emission-line region (Begelman et al.\ 1994), or dusty torus (Sikora
et al.\ 2009) by relativistic electrons, and high-energy particle
cascades (Mannheim et al.\ 1991).  A feature common to all models,
however, is that the $\gamma$-rays are associated with the compact
regions of relativistic jets energized by the central SMBHs.

\begin{figure}
\resizebox{\hsize}{!}{\includegraphics{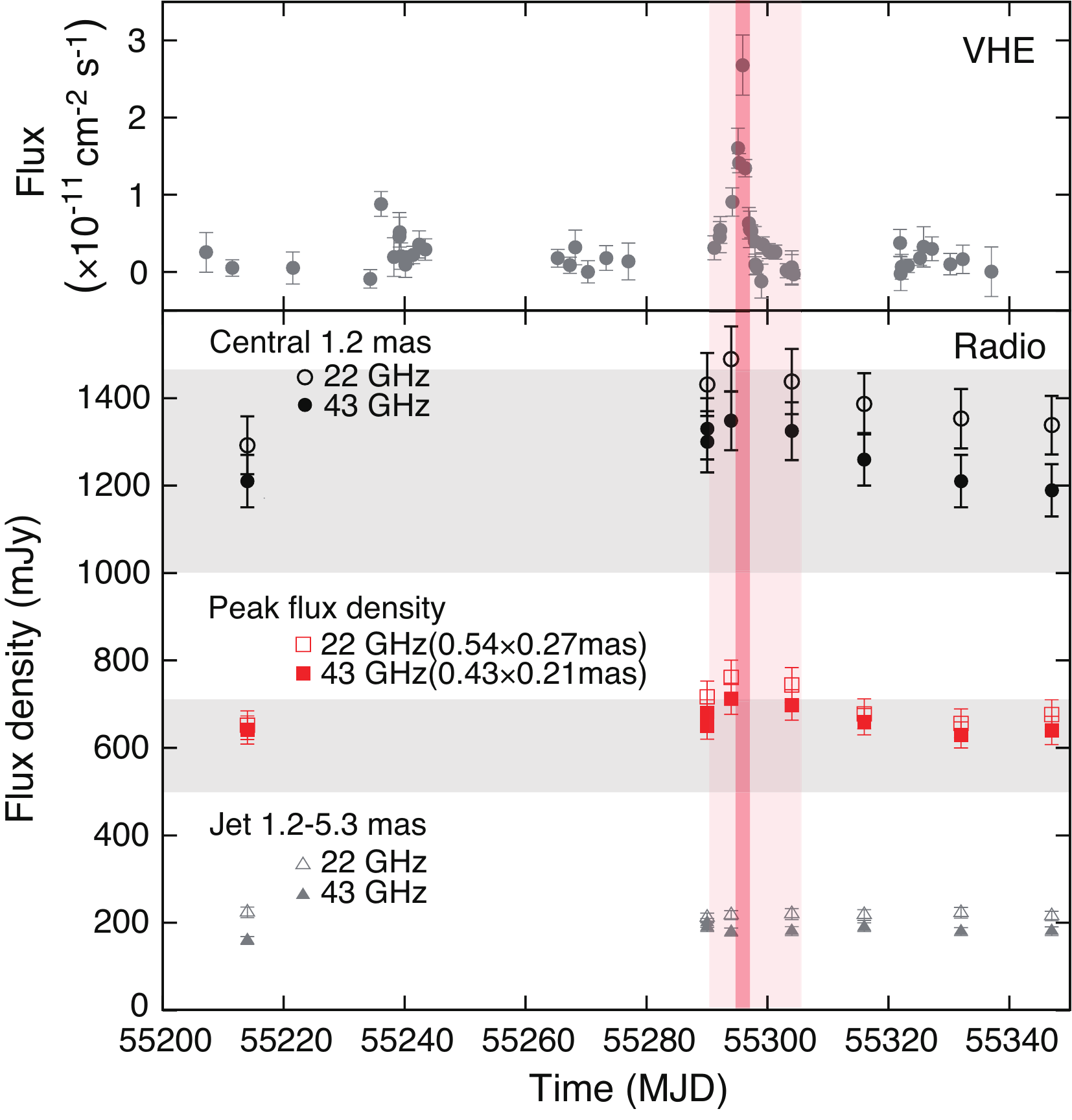}}
\caption{Centimeter light curves of the core of M87 correlate with the
  VHE $\gamma$-ray flare in 2010, suggesting that the VHE emission
  arises very near the black hole (Hada et al.\ 2012).
\label{fig-hada-m87}}
\end{figure}

High angular resolution observations from millimeter VLBI, combined
with $\gamma$-ray data from Fermi and ground-based Cherenkov
telescopes, offer opportunities to significantly increase our
understanding of VHE emission and particle acceleration mechanisms in
AGN.  The nearby radio galaxy M87 has exhibited several VHE flares in
recent years, with inconsistent correlation between the radio and
$\gamma$-ray photons (Hada et al.\ 2012, and references therein).  One
potential explanation is that variable magnetic field strengths near
the black hole cause variable opacity at centimeter wavelengths, with
the synchrotron self-absorption turnover frequency typically near
43~GHz.  Multifrequency monitoring using VLBI with phased ALMA during
a VHE flare would provide an observational test of this hypothesis
(Figure~\ref{fig-hada-m87}).

\begin{figure}
\resizebox{\hsize}{!}{\includegraphics{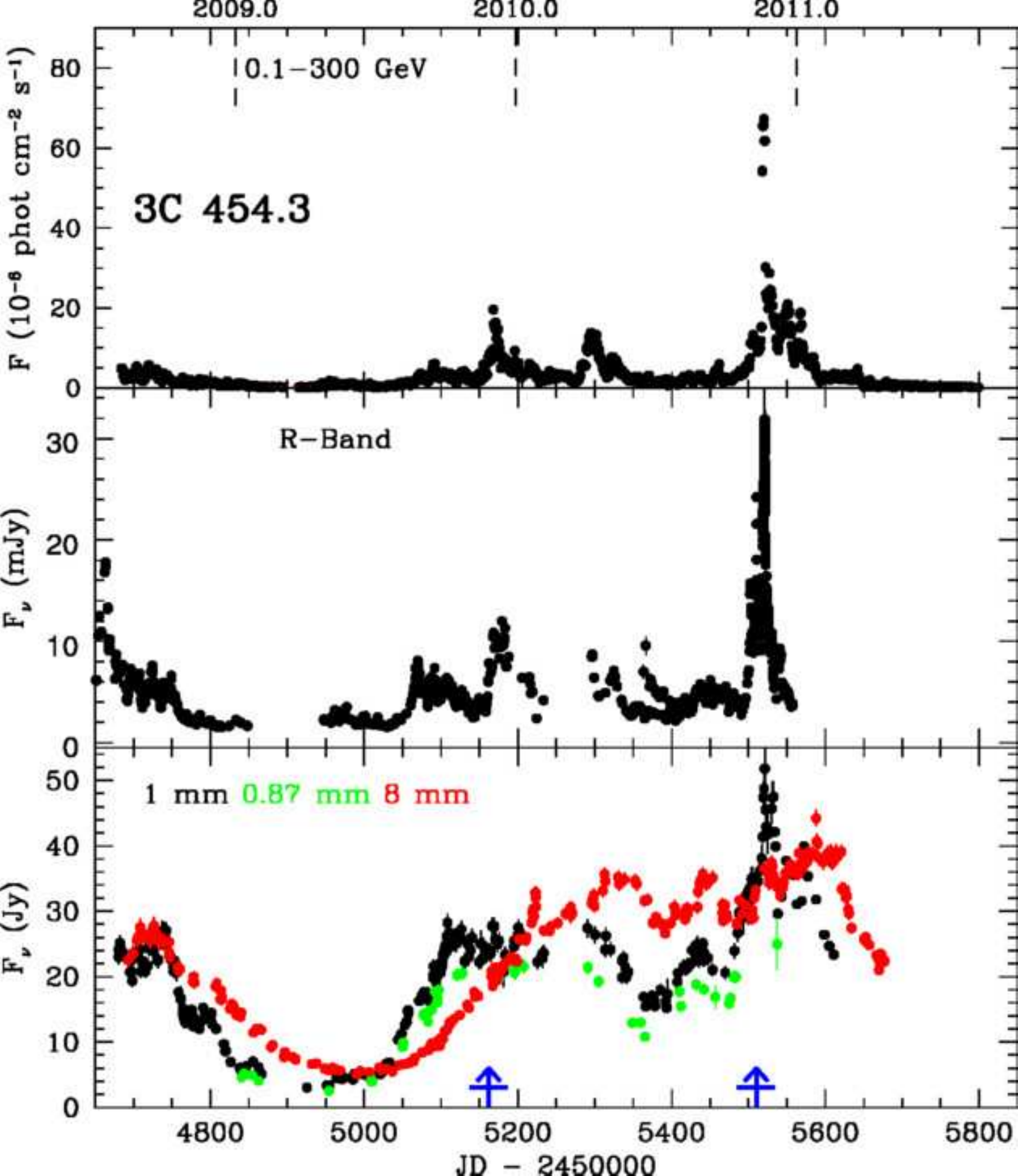}}
\caption{Multiwavelength light curves of the quasar 3C~454.3 during
  the extreme super outburst in late 2010.  The general behavior
  across wavebands is similar, but the detailed light curves are quite
  different.  The $\gamma$-ray light curve shows much greater correlation
  with the light curve at 1~mm than at 8~mm, suggesting that VLBI at
  submillimeter wavelengths may be better suited for probing the
  $\gamma$-ray emitting region.  Vertical blue arrows in the bottom panel
  indicate times when a new superluminal knot passed through the core
  in 7 mm VLBA images.  Data are from Jorstad et al.\ (in prep.) and
  Abdo et al.\ (2011).
\label{fig-3c454}}
\end{figure}

In blazars, high energy photons may be produced near and upstream of
th core.  The Fermi Large Area Telescope (LAT) instrument is gathering
well-sampled, continuous light curves of many bright $\gamma$-ray
blazars.  A collaboration led by A.\ Marscher (Boston U.) uses 7 and
3~mm VLBI to focus on a millimeter-bright subset (17) of these
objects.  So far, the results show that many $\gamma$-ray outbursts in
these objects occur when a new superluminal knot passes through the
core (Figure~\ref{fig-3c454}; Marscher et al.\ 2010; Jorstad et
al.\ 2010; Agudo et al.\ 2011a,b).  High-resolution imaging of this
region is therefore vital for understanding the mechanisms by which
these high-energy photons are produced (Blandford 2008).

\subsection{Double-Sided Radio Jets}

Cygnus~A, the closest ($z=0.057$) strong FR II radio galaxy, is a key
object for detailed studies of its prominent double-sided jet and
nucleus.  Relativistic beaming effects are reduced due to the large
inclination of the jet to the line of sight, allowing geometrical
effects to be disentangled from the internal jet speed.  This makes
Cyg~A an ideal candidate to study jet physics in an object similar to
those seen in more distant and luminous quasars (e.g., Barthel 1989).

The central engine and part of the innermost counterjet are likely
obscured by free-free absorbing material in a thick torus (Carilli et
al.\ 1994).  The absorbing material should be optically thin in the
(sub)millimeter regime.  Important physical parameters of the
obscuring torus (e.g., size, opacity, and temperature) can be
determined from the jet-to-counterjet ratio, which is
frequency-dependent.  A kinematic study of the two-sided jet structure
makes it possible to determine the jet speed and orientation in the
circumnuclear region (Krichbaum et al.\ 1998; Bach et al.\ 2002).  The
accuracy of the determination of this ratio requires knowledge of the
location of the true nucleus, located halfway between the footpoints
of the jet and counterjet.  This position can be determined more
accurately at higher observing frequencies due both to the lower
optical depth at high frequency and the higher angular resolution
provided by the observing array.  VLBI with ALMA in Bands 1 and 3 will
allow determination of the core position to within a small fraction of
the $\sim 100~\mu$as synthesized beamwidth.

Components of the main jet seem to accelerate from speeds of $\beta
\approx 0.1$-$0.2$ within 1~mas of the core (Krichbaum et al.\ 1998)
to $\beta \approx 0.3$-$0.7$ at 4~mas (Carilli et al.\ 1994; Bach et
al.\ 2002; Krichbaum et al.\ 2008).  VLBI observations at 7 mm reveal
significant structural variability in the counterjet, with a lower
velocity of $\beta \approx 0.1$-0.5.  The relatively low apparent
velocity in the counterjet could be due to an inherent asymmetry
between the two sides of the jet.  Jet stratification may be an
alternate explanation, if different layers of the counterjet are being
seen due to the viewing geometry.  High-fidelity imaging with phased
ALMA in Bands 1 and 3 as part of VLBI arrays will produce better
reconstructions of the jet emission, allowing observers to more
accurately model the structure in both the forward jet and counterjet
to determine whether the intrinsic structure is indeed asymmetric.

\subsection{Internal Jet Structure}

Models of BL Lac objects with TeV emission are self-inconsistent on
the Doppler factor of the jet, which Henri \& Saug\'{e} (2006) refer
to as the ``bulk Lorentz factor crisis of TeV blazars.''  In brief,
the two broad humps in the spectral energy distributions of these
objects are usually attributed to synchrotron self-Compton (SSC) and
inverse Compton processes associated with relativistic leptons.  The
fact that TeV photons are detected---and therefore not destroyed by
pair production processes---and the timescale of variability lead to
the conclusion that the bulk Lorentz factor must be very high (e.g.,
10 to 50) in the SSC region.  However, BL Lac objects are a small
subset of a larger class of FR I galaxies in which the jet is not
aligned with the line of sight.  The statistics imply that the bulk
Lorentz factor of the BL Lac subset should be close to 3, about an
order of magnitude lower than believed from SSC/TeV arguments.

\begin{figure}
\resizebox{\hsize}{!}{\includegraphics{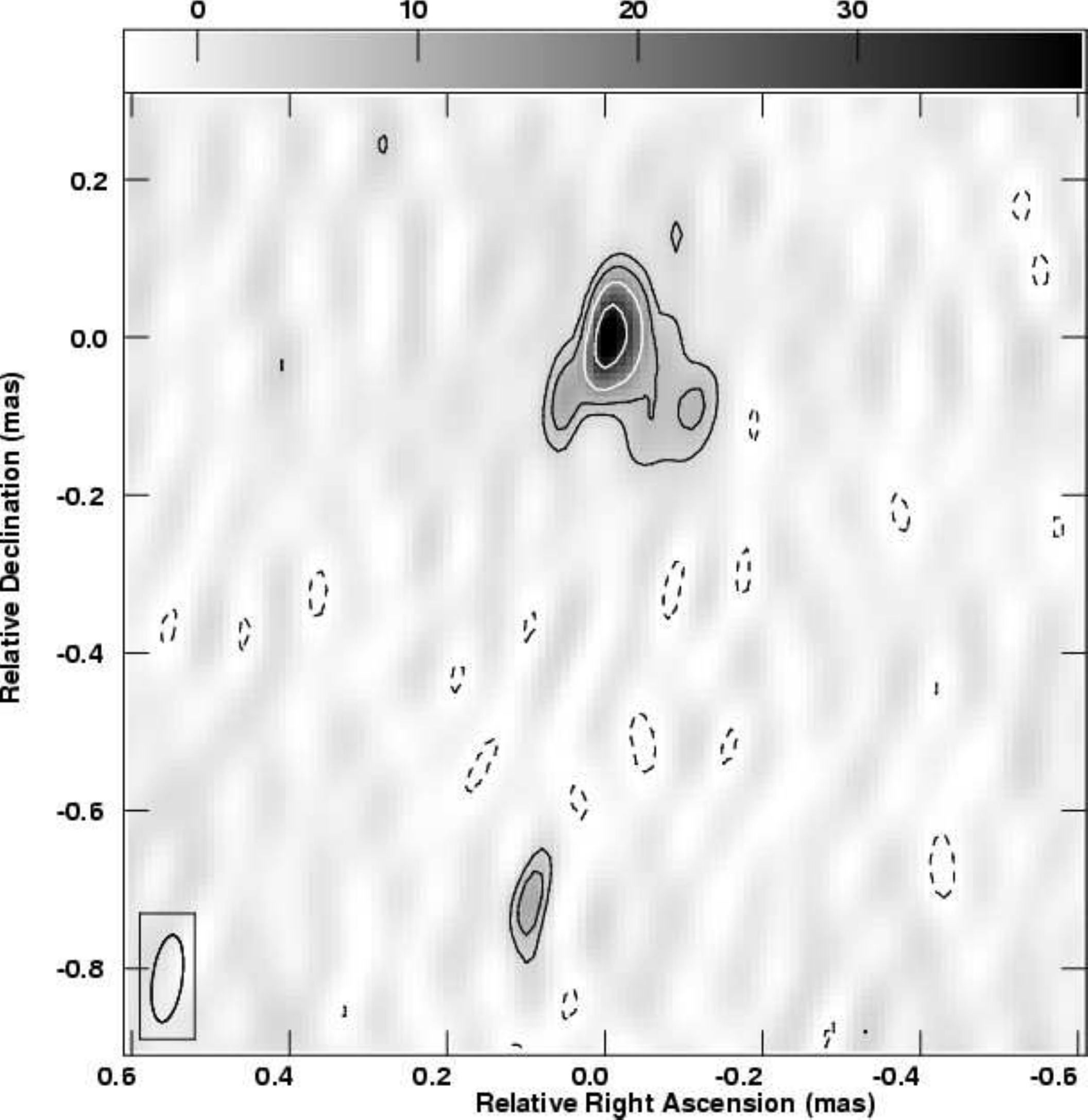}}
\caption{The 3 mm map of Mrk 501 (Giroletti et al.\ 2008) is
  suggestive of limb-brightened structure on scales of $\sim
  1000~r_\mathrm{Sch}$, as predicted by two-component jet models, but
  the image fidelity provided by the GMVA alone is not sufficient to
  be conclusive.
\label{fig-mrk501}}
\end{figure}

One possible resolution is if the jet is composed of two different
regions, a mildly relativistic jet containing and confining a pair
plasma.  This model predicts that the jet emission at 3~mm, where
opacity effects are less severe than at centimeter wavelengths, should
be limb-brightened at a distance of $\sim 1000~r_\mathrm{Sch}$ from
the black hole.  Polarimetric and spectral index observations support
the existence of spine/sheath structures in some jets (e.g., Attridge
et al.\ 1999; Aaron 1999; Edwards et al.\ 2000).  Typical jet sources
at nearby redshift ($z < 0.1$) are relatively weak, though, with flux
densities less than 1~Jy at 86 GHz.  In one epoch, GMVA observations
of the TeV BL Lac source Markarian 501 revealed what appears to be
limb-brightened jet structure on the relevant spatial scale
(Figure~\ref{fig-mrk501}; Giroletti et al.\ 2008).  However, the image
fidelity is poor, and later observations in good observing conditions
were unable to confirm the result (Koyama et al., in prep.).  This
indicates that existing instrumentation may just barely be able to
reveal features of great interest, but improvements in sensitivity and
image fidelity, as would be provided by including phased ALMA in Band
3 VLBI observations, are necessary for a real breakthrough.

\subsection{Polarization and Magnetic Fields}

Full-polarization VLBI observations at high frequencies have important
advantages in the study of AGN jets.  Faraday rotation and
depolarization effects are reduced at higher frequencies.  Due to
synchrotron self-absorption, the observed emission at millimeter
wavelengths comes from closer to the jet base.  The polarized emission
tracks the magnetic field, which plays a crucial role in jet formation
and collimation, from the neighborhood of the black hole up to
extragalactic distances (e.g., Broderick \& Loeb 2009; Homan et
al.\ 2009; G\'{o}mez et al.\ 2011; O'Sullivan et al.\ 2011;
Tchekhovskoy et al.\ 2011; McKinney et al.\ 2012).  Polarimetric VLBI
observations can probe the strength (e.g., Savolainen et al.\ 2008;
O'Sullivan \& Gabuzda 2009) and spatial structure of the magnetic
fields and can therefore be used to test theories of jet
structure. For instance, if the spine of the jet is pressure dominated
but confinement is provided by toroidal magnetic fields, the gradient
of Faraday rotation across the jet should have a constant sign along
the jet direction (Blandford 2008).  Indeed, such rotation measure
gradients have been detected in several jets (e.g., Asada et
al.\ 2002; Gabuzda et al.\ 2004; G\'{o}mez et al.\ 2008; Hovatta et
al.\ 2012) and interpreted as indicating the presence of a toroidal or
helical magnetic field in the jet sheath.  However, the region in
which the rotation measure gradient is detected is located at more
than $10^5~r_\mathrm{Sch}$ from the black hole, beyond the
acceleration and collimation zone.  High angular resolution
observations with high sensitivity and image fidelity would allow
similar results to be obtained nearer to the central black hole.

\begin{figure}
\resizebox{\hsize}{!}{\includegraphics{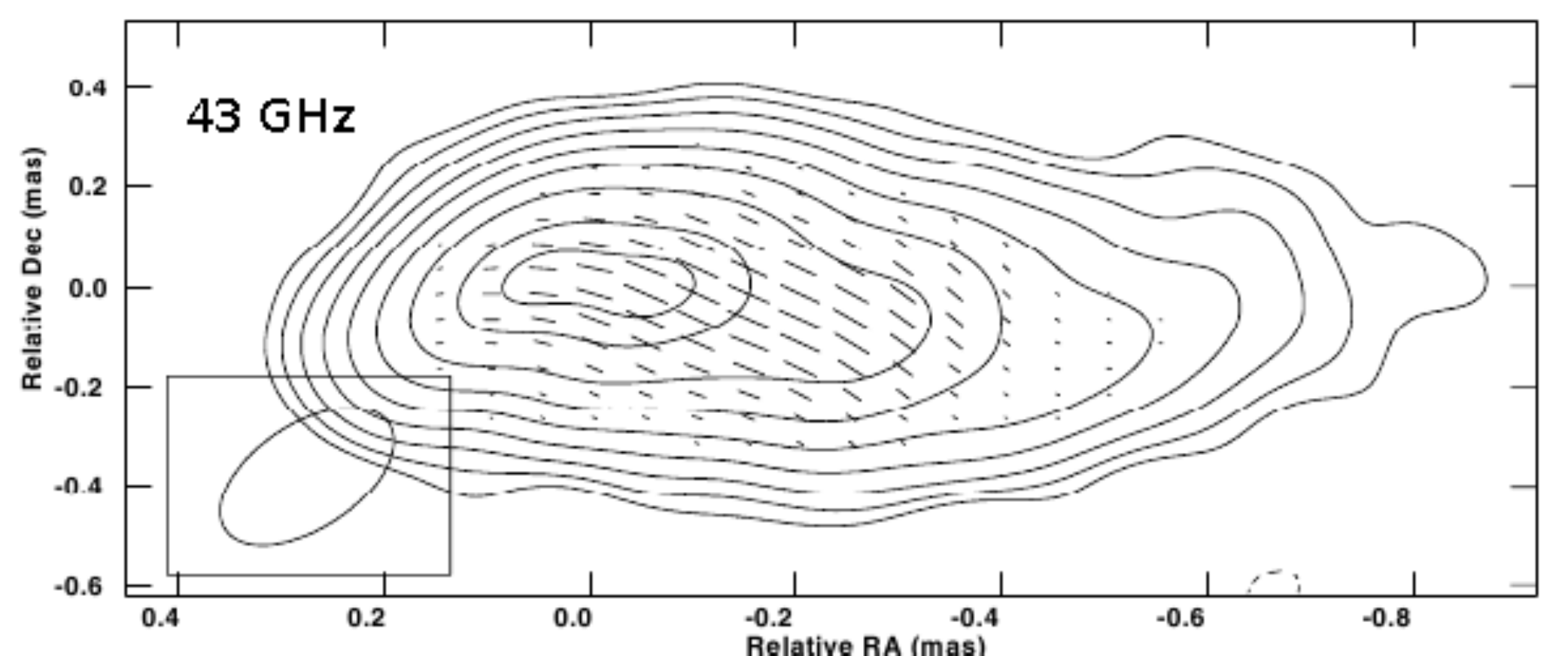}}\\
\resizebox{\hsize}{!}{\includegraphics{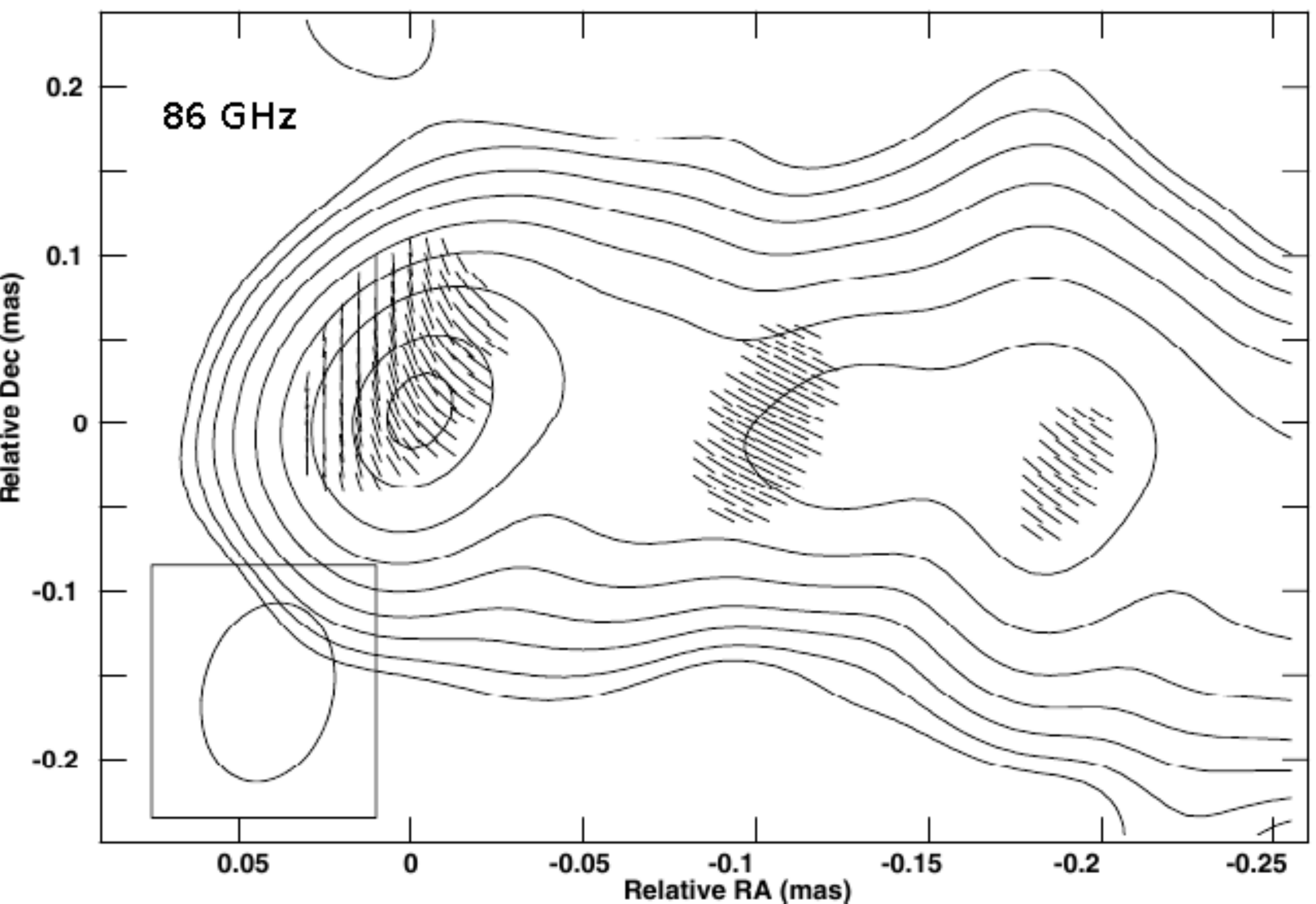}}
\caption{Polarization maps of 3C~345 at 43~GHz (Jorstad et al.,
  in prep.) and 86~GHz (Mart\'{\i}-Vidal et al.\ 2012).  The
  very compact polarized region close to the core at 86 GHz, whose
  electric vectors are rotated by almost 90 degrees with respect to
  the rest of the jet, is not detectable at 43~GHz, possibly due in
  part to beam depolarization.
\label{fig-3c345}}
\end{figure}

The inclusion of phased ALMA will dramatically improve polarization
capability at millimeter wavelengths.  The quality and purity of the
polarization signal in the ALMA receivers will allow dramatic
increases in the fidelity of full-polarization VLBI images at high
frequencies, which can be the limiting factor for polarimetric imaging
at 86 GHz; for example, the noise cutoff can exceed 50\% of the
polarization peak in current GMVA observations (Mart\'{\i}-Vidal et
al.\ 2012).  The large collecting area is also critical for improving
the sensitivity of millimeter VLBI arrays, since the polarized flux is
generally only a small fraction (often 10\% or less) of the total
intensity signal.  The higher angular resolution available on long
baselines to ALMA is important for reducing beam depolarization, which
reduces sensitivity to structures that vary in polarization position
angle on scales comparable to the beam size (Figure~\ref{fig-3c345}).

\subsection{Jet Precession and Binary Black Holes}

Many AGN jets are observed to exhibit swings in the position angle of
the inner jet (e.g., Stirling et al.\ 2003; Tateyama \& Kingham 2004;
Agudo et al.\ 2007).  Multiple hypotheses have been advanced to
explain jet precession, including binary black hole systems (Begelman
et al.\ 1980), accretion disk instabilities, and the Bardeen-Petterson
effect (Caproni et al.\ 2004 and references therein).  Progress on
understanding which mechanisms apply to specific cases has been
unsteady.  For example, the periodic regular outbursts in OJ~287 have
been explained as arising due to interactions of a binary pair of
SMBHs (Sillanp\"{a}\"{a} et al.\ 1988).  Although this model has been
developed and refined over the past two decades, no model is
consistent with the totality of the observational evidence, including
the timing of the outbursts, radio imaging, and optical polarimetry
(Villforth et al.\ 2010).  In some sources, there is no consensus even
as to whether or not the jet precession is periodic (e.g., Mutel \&
Denn 2005).  Regular monitoring with high angular resolution and high
sensitivity is required to clarify the observational picture, a
necessary prerequisite for interpretation.  The mechanisms producing
precession in binary AGN systems with accretion disks may also have
applicability to Galactic microquasars (e.g., Katz 1997).

Supermassive binary black holes, which form during major galactic
mergers, are predicted to evolve to a close pair and subsequently
coalesce.  Clear evidence exists for nearby interacting galaxies with
jets as well as for black holes with separations on scales from
kiloparsecs down to parsecs (Owen et al.\ 1985; Komossa et al.\ 2003;
Rodriguez et al.\ 2006; see also review in Komossa 2006).  Binary
black hole systems may spend a large fraction of their lifetime at
subparsec scales (Begelman et al.\ 1980), corresponding to angular
scales of hundreds of microarcseconds at typical distances
(Kudryavtseva 2008).  The long baselines and sensitivity of phased
ALMA are necessary to study the interactions within these close binary
systems, such as the strength of the radio emission from close black
hole pairs and the presence of a secondary jet.  The presence of a
second black hole causes periodic or quasi-periodic variability of the
radio flux density and VLBI core flux (e.g., Qian et al.\ 2007,
Kudryavtseva et al.\ 2011).  Astrometry of the innermost jet of these
candidate binary systems willl test whether the black holes follow
elliptical trajectories around a common mass center, providing the
first direct evidence that these systems do indeed contain more than
one supermassive black hole.

\subsection{Microquasars}

Microquasars have emerged as very important local analogues of
AGN. They represent an opportunity to observe black hole dynamics and
jet formation on a much smaller scale. ALMA will open up an important
new way to study these because the field is currently limited by the
need to go to higher frequencies for better resolution and
sensitivity.

Microquasars are the stellar remnant of a star that has collapsed to
form a compact object, such as a neutron star or black hole, and has
remained gravitationally bound to its binary companion star.  Matter
is transferred from the companion star to the compact object either
via a stellar wind (for high-mass stars) or a Roche lobe (for low-mass
stars).  An accretion disk forms around the compact object, releasing
extremely high energies within a very small volume.  Microquasars
release a small fraction of the inflow matter by accelerating
particles into highly collimated bipolar jets; hence they have a
similar morphology to quasars, but on much smaller scales.

\begin{figure}
\resizebox{\hsize}{!}{\includegraphics{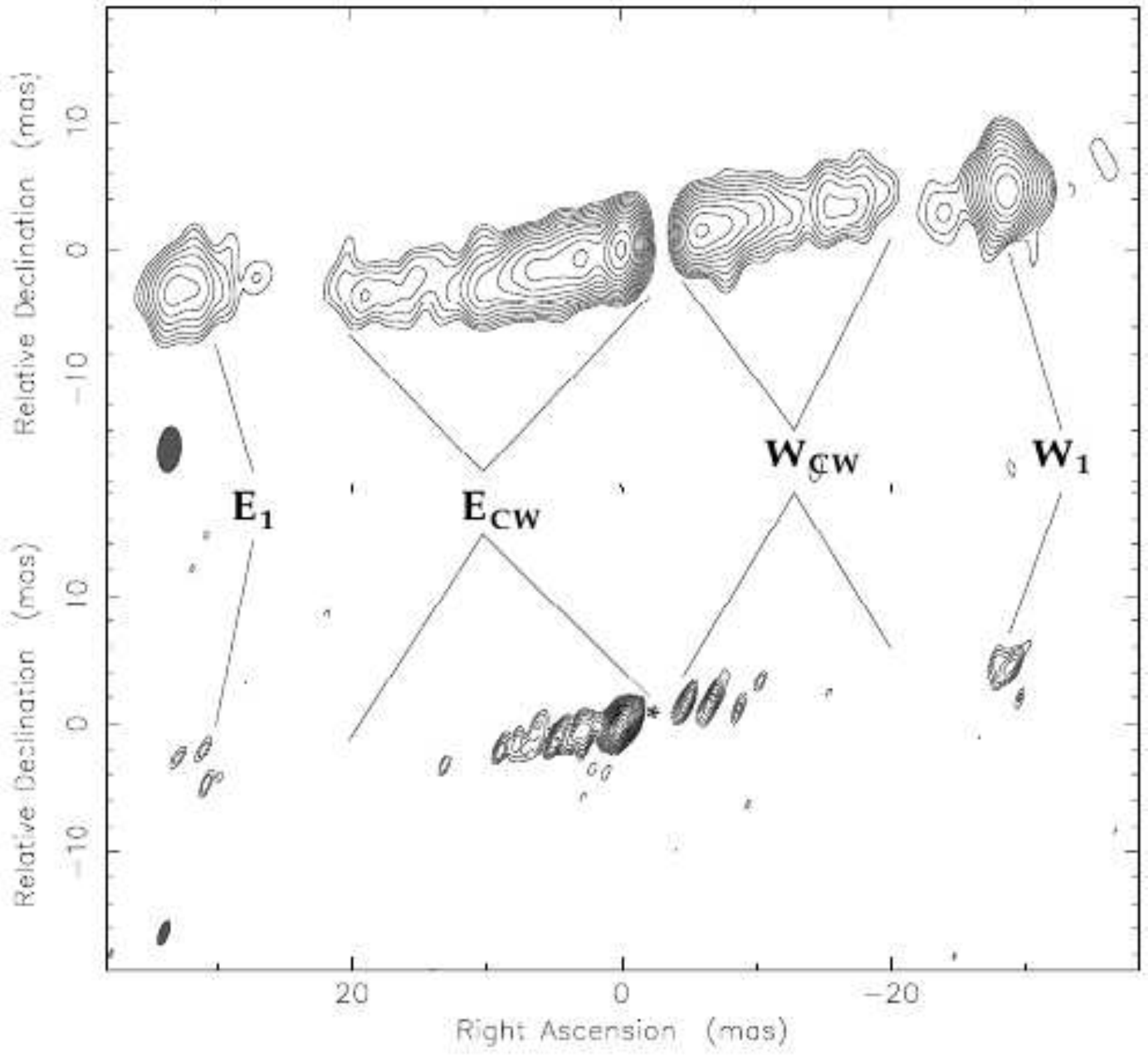}}
\caption{Images of the jet in SS433 at 5 GHz (upper image) and 15
GHz (lower image), from Paragi et al.\ (1999).
\label{fig-ss433}}
\end{figure}

Like AGNs, the jets from microquasars have a high luminosity,
nonthermal spectra, and polarization properties consistent with
self-absorbed synchrotron radiation (Hjellming \& Johnston 1988).  The
minimum size of the jet region is related to the mass of the compact
object and the turnover frequency at which the emission transitions
from being self-absorbed to optically thin: $\nu_\mathrm{max} \propto
r_\mathrm{min}^{-1} \propto M_\mathrm{BH/NS}^{-1}$.  Therefore,
emission from the compact jet can dominate the SED from stellar-mass
sources at much higher frequency than in the cores of AGN (e.g.,
SS433; Figure~\ref{fig-ss433}). While we know that shocked ejecta from
these sources can have superluminal apparent motion (Dhawan et
al.\ 2000), the velocity of the compact jet is still unknown.  We do
not know exactly where superluminal shocks are formed or how their
formation time is associated with changes in the accretion disk.

One of the most persistently bright Galactic microquasars is GRS
1915+105, which appears to be in near constant outburst.  It is
estimated that the source has a relatively high accretion rate ($\sim
0.1~L_\mathrm{Edd}$), which feeds a jet moving at up to 0.98c (Mirabel
\& Rodriguez 1994; Fender et al.\ 1999), giving an overall spectrum
that is flat and bright ($> 100$~mJy) from radio to submillimeter
wavelengths.  Moreover, the source has one of the largest accretion
disks in the Galaxy.  With an estimated outer radius of $10^{12}$~cm
(Done et al.\ 2004), the disk size corresponds to an angular scale of
$\sim 10~\mu$as, assuming a distance of 10 kpc.  This matches very
well the angular resolution of VLBI baselines to phased ALMA at
millimeter and submillimeter wavelengths.  VLBI with phased ALMA will
be able to directly resolve the accretion disk and image the region
where the particles are accelerated to superluminal apparent
velocities.  It will be possible to measure the expansion of the
shocked jets, measure the tangential jet cross-section, and pinpoint
the position of the central black hole or neutron star.

ALMA will also be able to study the proposed free-free absorption
region and observe anisotropy in the stellar wind around high-mass
stars such as Cygnus X-1 (Brocksopp et al.\ 2002).  Located at a
distance of 1.86~kpc (Reid et al.\ 2011), the black hole X-ray binary
Cyg~X-1 is observed to transition between low/hard and high/soft X-ray
states.  It is theorized that these states are characterized by the
existence of a thick advection dominated accretion flow (ADAF) inside
a transition radius and a geometrically thin disk with hot corona
outside this radius (Narayan 1996).  The low/hard and high/soft states
may represent the two extremes of this picture, with the system
transitioning from an ADAF into a thin disk as the accretion rate
increases (Esin et al.\ 1998).  Simultaneous radio and X-ray
monitoring suggests that the disk, corona, and jet are all coupled,
with the location of the jet base around $20~GM/c^2$ above the black
hole rather than near the inner edge of the accretion disk (Miller et
al.\ 2012).  Cyg~X-1 is seen to produce a radio jet extending out to
about 15~mas (e.g., Stirling et al.\ 2001), but, unexpectedly, VLBI
observations at centimeter wavelengths do not provide any evidence for
the formation of knots or shocks as Cyg X-1 transitions between the
low/hard and high/soft states (Rushton et al.\ 2012).  Nevertheless,
some differences must exist between the jet in these two states, since
the total radio flux is modulated by the orbital period (5.6 days;
Pooley et al.\ 1999) in the low/hard state but not in the high/soft
state (Rushton et al.\ 2012); furthermore, the millimeter flux density
is substantially higher in the high/soft state than in the low/hard
state (Tigelaar et al.\ 2004).  It is possible that structural changes
in the jet are being masked by the high optical depth near the radio
core at centimeter wavelengths, although the very flat spectrum from
radio through millimeter wavelengths suggests that a combination of
self-absorbed synchrotron, free-free, and thermal emission combine to
produce the emission from Cyg~X-1 (Fender et al.\ 2000).
High-resolution VLBI observations in Bands 1, 3, and 6 including
phased ALMA will have a much greater capability of detecting jet
changes during the transition between the hard and soft states.  Due
to their rapid timescale of variability, microquasars such as Cyg X-1
represent unique probes of structural changes in small-scale
disk-corona-jet systems, but the same physical processes may be
applicable to AGN/jet systems as well, in which variability due to
changes in the accretion rate occur much more slowly.

Finally, the inclusion of ALMA in VLBI arrays will enable
high-precision astrometry (\S~\ref{astrometry}) of X-ray binaries.
Parallax measurements provide a direct geometric distance to the
source, an important prerequisite for extracting the masses of the
objects in these systems (see, e.g., Orosz et al.\ 2011).  Detections
of the orbital motion of the jet base place strong constraints on
physical properties of the binary system, including estimates of the
semimajor axis of the orbit and the ratios of the masses in the
system.  Cyg~X-1 serves as a good example of the power of these
techniques (Reid et al.\ 2011; Orosz et al.\ 2011), but the size of
the orbital motion in angular units ($\sim 70$~$\mu$as) is larger than
most other X-ray binaries.  The extra resolution provide by VLBI with
phased ALMA in Bands 1 and 3 may be necessary for other sources.

\section{Pulsars}\label{pulsars}

Pulsars are broadband, steep-spectrum radio sources.  Pulsars are
typically observed at frequencies of a few GHz because the timing
precision scales with S/N while the effects of the interstellar medium
become less severe at higher frequencies (e.g., Lorimer \& Kramer
2005).  A number of pulsars are strong enough for individual pulses to
be detected even at frequencies above 10 GHz (Xilouris et
al.\ 1994), providing clues for identifying the as-yet unknown origin
of the coherent emission process of pulsars (Lorimer \& Kramer 2005).
It is the quest for solving this ``pulsar emission mystery'' that is
one of the main motivations for pulsar observations at very high
frequencies.  The highest frequency at which a radio pulsar (PSR
B0355+54) has been observed to date is~87 GHz, using the IRAM 30-m
telescope (Morris et al.\ 1997).  Being able to study pulsars at
millimeter wavelengths is also important for our understanding of the
relationship between neutron star populations and for the search for
pulsars in the Galactic Center region.  Detecting a pulsar in orbit
around Sgr~A* would allow astronomers to measure the properties of the
central SMBH with unprecedented precision.

\subsection{Deciphering the Emission Mechanism}

A number of pulsars have been detected at millimeter wavelengths: nine
sources at 32~GHz, four at 43~GHz, and one at 87~GHz (see L\"{o}hmer
et al.\ 2008 for a recent summary).  A sensitive millimeter telescope
is needed to increase this number.  The location of ALMA is especially
appropriate, since 72\% of all known pulsars are located in the
Southern hemisphere.

\begin{figure}
\resizebox{\hsize}{!}{\includegraphics{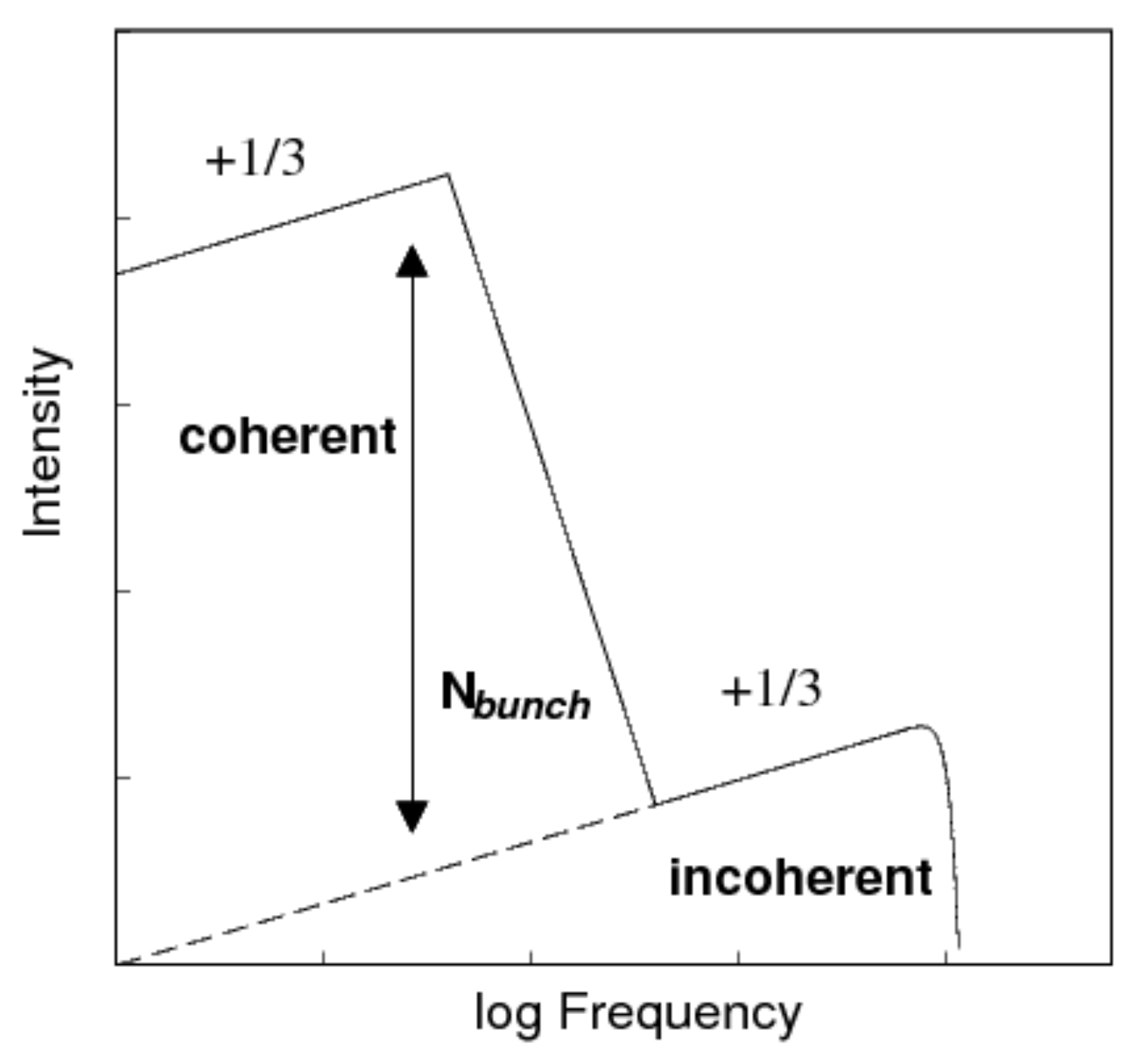}}
\caption{Idealized spectrum expected for curvature emission from
  bunched particles (adapted from Michel 1982, 1991) as an example of
  the transition from coherent to incoherent emission (Kramer 1995).
  In this case, the coherent radiation is enhanced with respect to the
  incoherent emission by a factor corresponding to the number of
  particles within a bunch.
\label{fig-pulsarcoherence}}
\end{figure}

Previous observations indicate that the emission properties at these
very high frequencies are particularly interesting for understanding
the creation of coherent pulsar emission.  Two out of the nine sources
detected above 30 GHz show a peculiar change in the flux density,
indicating a flattening or even an increase in the flux density with
increasing frequency (Kramer et al.\ 1996; L\"{o}hmer et al.\ 2008).
This change in the spectrum was not unexpected, since it has been
known for a long time that the Crab pulsar's infrared flux density is
much higher than the high-frequency radio flux density (Lyne \& Smith
2005).  A similar result comes from observations of the Vela pulsar
(Danilenko et al.\ 2011).  One natural interpretation for a spectral
change in the range between centimeter and infrared observations is
that the coherence conditions for normal pulsar radio emission break
down and an incoherent component of the emission takes over (Kramer
1995; Michel 1982, 1991).  Indeed, many of the poorly-understood
(though well-studied) striking properties of the radio emission are
likely a consequence of the coherence process, including
nanosecond-scale pulse structure, very high brightness temperatures
exceeding $10^{36}$~K, linear polarization fractions approaching 100\%
(often with a strong circularly-polarized component), and steep radio
spectra with spectral indices ranging between $0$ and $-3.4$.  As
plasma moves along curved magnetic field lines in the magnetosphere,
there should always be an underlying incoherent component that could
become detectable once the coherence condition breaks down
(Figure~\ref{fig-pulsarcoherence}).

The intrinsic coherence length could be determined by detecting this
break in the spectrum, providing invaluable clues to understanding the
emission process.  Phased ALMA would allow the first systematic study
of pulsars at frequencies of 90~GHz or higher.  With 50 antennas,
phased ALMA could reach an equivalent continuum flux density of
$8.8~\mu$Jy in 1~hr at 92~GHz, detecting PSR B0355+54 easily with S/N
$\approx 60$ and PSR B1929+10, one of the two pulsars with indications
for a spectral turn-up at high frequency, with S/N $> 20$.  Assuming
an average spectral index of $-1.6$, 3\% of all known pulsars could be
studied with one hour of integration, and almost 10\% with a five-hour
integration time.  This would result in the detection of nearly 100
pulsars, providing excellent statistics for the study of emission
processes at ALMA Band~3 frequencies.

\subsection{Magnetars and Their Relationship to Pulsars}

Magnetars are a special class of neutron stars that typically have
rotation periods between 5 and 12~s.  They get their name from their
very large inferred magnetic fields (in excess of $10^{14}$~G), which
appear to power these sources during their high-energy outbursts.
Their radio brightness seems to be a transient event that may be
related to or even triggered by the high-energy outburst (Camilo
2008).  When radio-bright, magnetars show similarities to pulsars but
retain important differences: they are very highly linearly polarized,
show an unusually large degree of variation in their pulse shape, and
have a very flat and variable flux density spectrum (Camilo et
al.\ 2008; Kramer et al.\ 2007).  The magnetar XTE J1810-197 has been
detected with the IRAM 30-m telescope at a frequency as high as 144~GHz
with a flux density of about 1.2~mJy (Camilo et al.\ 2007).  Studies
of magnetar properties would therefore be easily possible with phased
ALMA and should allow comparison of magnetar and radio pulsar
properties, advancing understanding of radio-emitting neutron stars in
general.  In particular, there are questions of why the radio spectrum
of magnetars is so flat and to what frequencies it could extend.

\subsection{Probing the Spacetime Around Sgr~A*}

Pulsars in orbit around Sgr~A* will be superb probes to study the
properties of the central SMBH (e.g., Wex \& Kopeikin 1999; Liu et
al.\ 2012).  Timing observations of such pulsars complement VLBI
imaging for testing general relativity because the two techniques are
based on different and independent observables.  Finding a normal,
slowly-rotating pulsar in a reasonable orbit around Sgr~A* could under
ideal circumstances measure the mass of the black hole and test the
cosmic censorship conjecture and no-hair theorems to high precision
(Liu et al.\ 2012).  Finding pulsars in the Galactic Center will
provide invaluable information about the Galactic Center region
itself.  The characteristic age distribution of the discovered pulsars
will give insight into the star formation history.  Millisecond
pulsars can be used as accelerometers to probe the local gravitational
potential.  The dispersion and scattering measures and their
variability can serve as probes of the distribution, clumpiness, and
other properties of the central interstellar medium.  Faraday rotation
can provide measurements of the central magnetic field.

Given the huge rewards for finding and timing pulsars in the Galactic
Center, various efforts have been conducted to survey the inner Galaxy
(Kramer et al.\ 2000, Johnston et al.\ 2006; Deneva et al.\ 2009;
Macquart et al.\ 2010).  None of these efforts has been successful,
despite the expectation to find more than 1000 pulsars in the Galactic
Center region, including millisecond pulsars (e.g., Wharton et
al.\ 2012) and highly eccentric systems including both a stellar black
hole and a millisecond pulsar (Faucher-Gigu\`{e}re \& Loeb 2011).
However, finding these pulsars is incredibly difficult due to the
severely increased interstellar scattering produced by the highly
turbulent medium in the vicinity.  Scattering leads to pulse
broadening that cannot be removed by instrumental means and that
renders the source undetectable as a pulsar, in particular if the
scattering time exceeds a pulse period.  The scattering time is a
strong function of frequency (Lorimer \& Kramer 2005), so the
aforementioned pulsar searches have been conducted at ever-increasing
frequencies---the latest being conducted around 20~GHz.  The
difficulty in finding pulsars at these frequencies is two-fold.
First, the flux density is significantly reduced due to the steep
spectra of typical pulsars.  Second, the reduced dispersion delay,
which needs to be removed but also acts as a natural discriminator
between pulsar signals and terrestrial interference, makes
verification of real signals difficult.

\begin{figure}
\resizebox{\hsize}{!}{\includegraphics{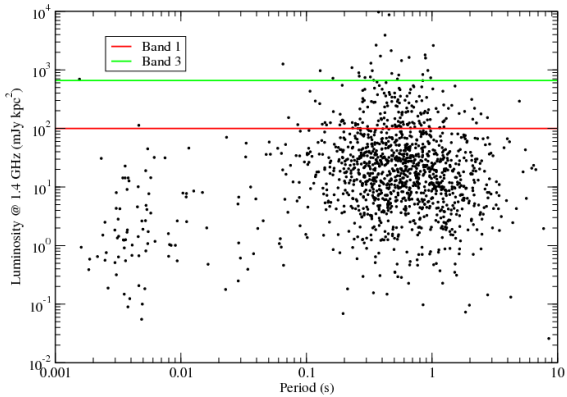}}
\caption{Luminosity of the known radio pulsar population.  The survey
  sensitivity of searches with ALMA Bands 1 and 3 are indicated with
  red and green lines, respectively.  These limits were derived using
  a distance of 8.5~kpc to the Galactic Center and assuming an average
  spectral index of $-1.6$.
\label{fig-pulsarluminosity}}
\end{figure}

Recording baseband data with the ALMA beamformer will help in both
respects.  ALMA's sensitivity and location in the Southern hemisphere
will enable deep pulsar searches of the Galactic Center.  The baseband
data can be used to study the frequency structure of the signal in
great detail, for instance, by synthesizing a very fine polyphase
filterbank or searching directly for chirped signals.  The scattering
and dispersion times are low enough at ALMA frequencies that it will
be possible to detect even millisecond pulsars at Band~3. Scaling from
the properties of known pulsars, a search with 50 phased ALMA dishes
would detect pulsars with a 1.4~GHz luminosity of about 100
mJy\,kpc$^2$ at Band~1 and 660 mJy\,kpc$^2$ at Band~3
(Figure~\ref{fig-pulsarluminosity}).  High-frequency pulsar searches
could be done commensally with other observations.  For instance, the
same data taken for Band~6 VLBI observations or routine ALMA Band~3
observations of the Galactic Center could be used for pulsar searches,
thus leveraging the same ALMA observing time for two different
scientific purposes.  Indeed, the recent serendipitous detection of a
young pulsar in the vicinity of the Galactic Center (Eatough et
al.\ 2013) provides evidence in support of a yet-undetected population
of pulsars near Sgr~A*.

\section{Gravitational Lens Systems and Absorption at Cosmological
  Distances}\label{absorbers}

\subsection{Dark Matter}

Foreground galaxies may gravitationally lens high-redshift galaxies or
AGN into multiple images.  If small-scale overdensities are present
within the dark matter halo of the lensing galaxy, secondary
magnifications and distortions of the separate images can also occur
(see review in Zackrisson \& Riehm 2010).  Whereas morphological
features intrinsic to the source will be present in all images,
lensing due to dark halo substructure will produce uncorrelated
effects in the images (e.g., Inoue \& Chiba 2005).  Searches for such
small-scale lensing effects can probe the dark subhalos predicted by
standard cold dark matter (CDM) simulations or more compact forms of
halo substructure produced in alternative structure formation
scenarios, such as primordial black holes or ultracompact minihalos.
This is particularly interesting in the dwarf-galaxy mass range, where
the subhalos seen in CDM simulations greatly outnumber luminous
satellite galaxies, suggesting that a vast population of extremely
faint or completely dark subhalos may be awaiting detection in the
halo of every large galaxy.  The superior resolution and sensitivity
provided by VLBI with phased ALMA will constrain halo substructure at
lower masses than previously possible.  Observations of lensed AGN
jets in Band 3 would allow for the detection of black holes with
masses of $\sim 10^3$--$10^6~M_\sun$ or ultracompact minihalos with
masses of $\sim 10^6$--$10^8~M_\sun$ (Zackrisson et al.\ 2013).
Strongly lensed submillimeter galaxies can, due to their larger
intrinsic sizes ($\sim 100$~pc; Swinbank et al.\ 2010), be used to
probe lower-density substructures, and observations in Band 7 could
make standard CDM subhalos detectable down to at least $\sim
10^7~M_\sun$.

\subsection{Absorption}

\begin{figure*}
\resizebox{0.25\hsize}{!}{\includegraphics{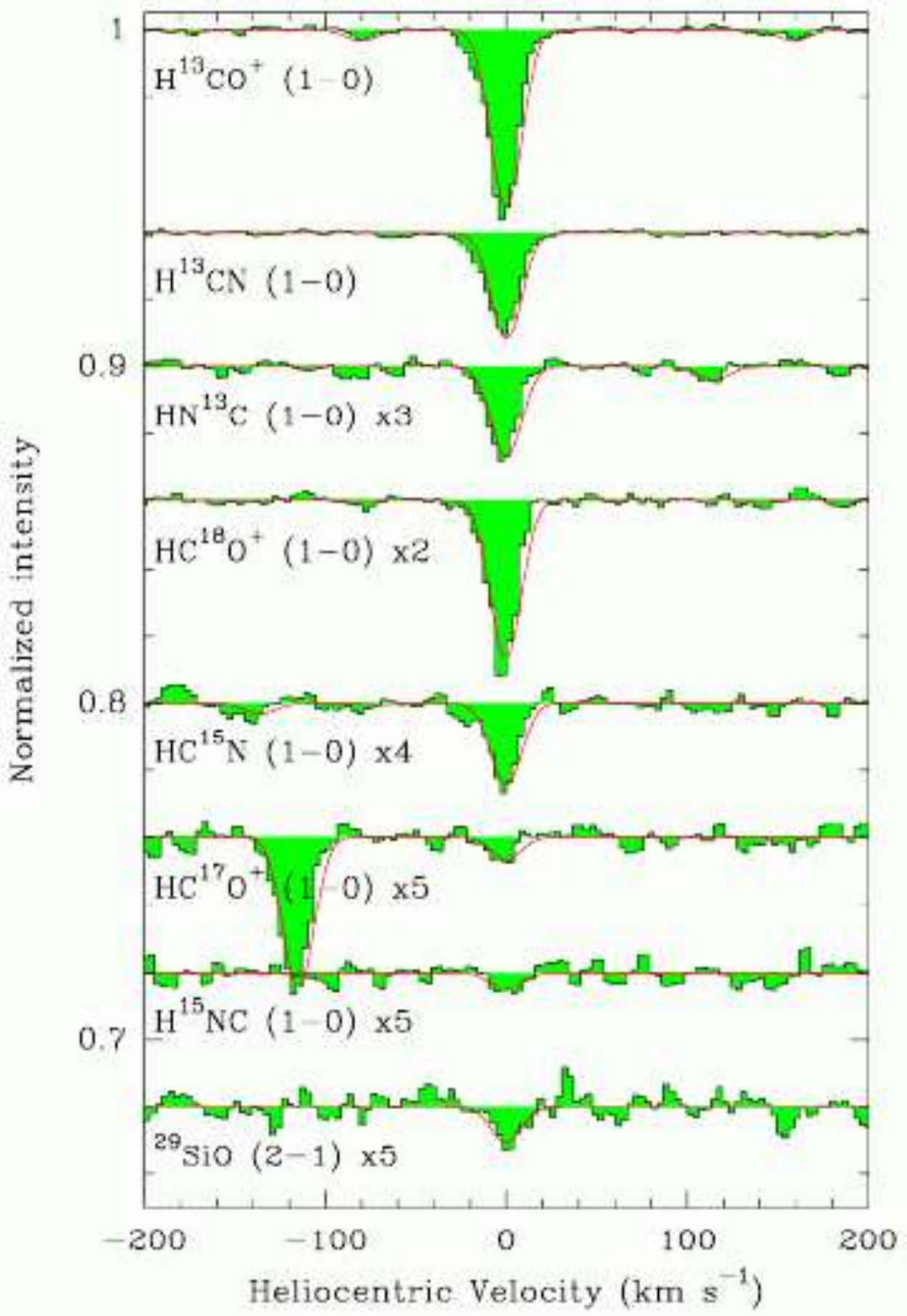}}
\resizebox{0.74\hsize}{!}{\includegraphics{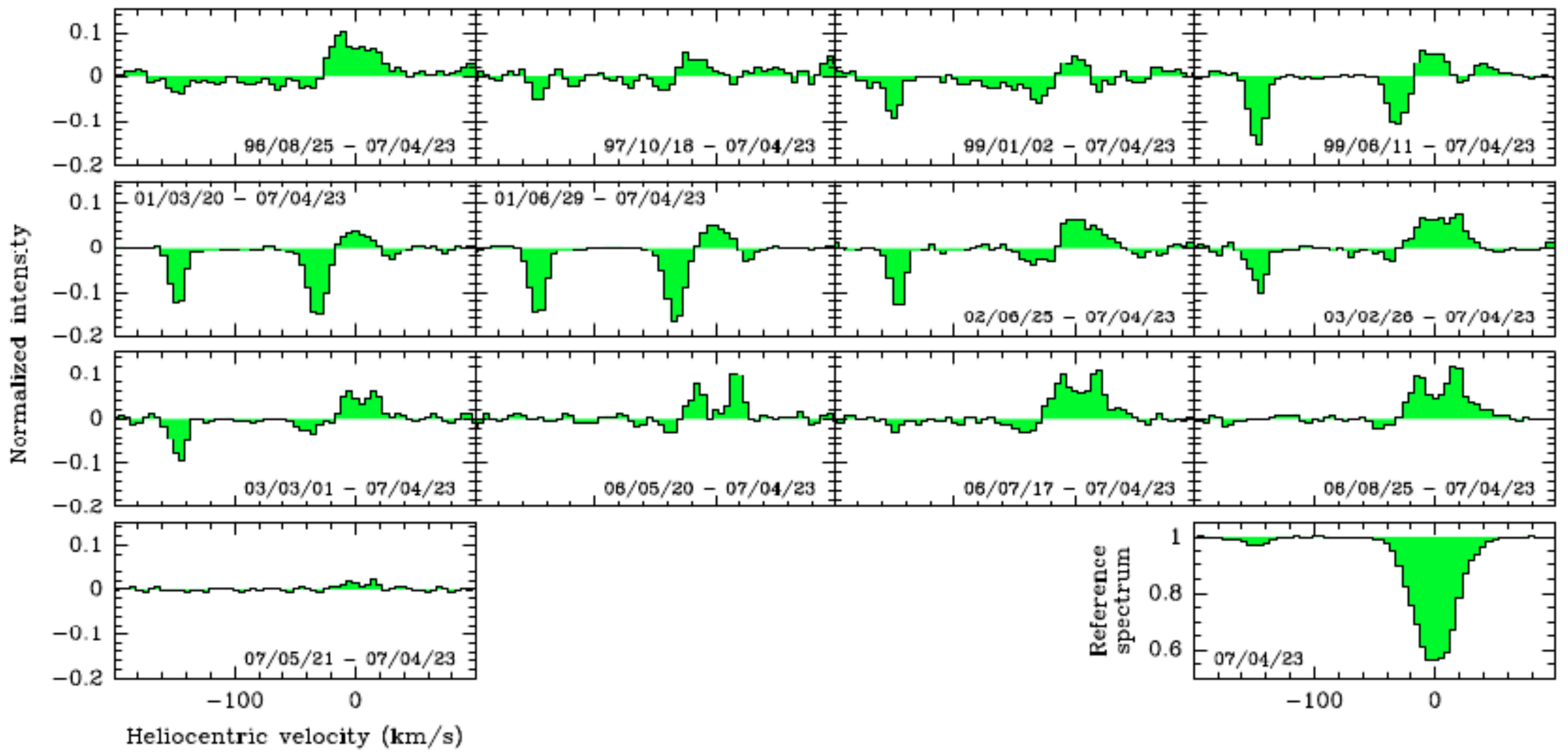}}
\caption{Molecular absorption in the absorbing system toward PKS
  1830$-$211.  Many molecular species and isotopologues are detected
  in the millimeter regime in this system (left, Muller et al.\ 2011).
  Their variability indicates the the absorbing clouds are very
  compact (right, Muller \& Gu\'{e}lin 2008).
\label{fig-absorbers}}
\end{figure*}

Most potential VLBI targets require very high brightness temperatures
to be detected.  In the case of spectral-line VLBI, maser
amplification is required to produce sources that can be detected
through their own emission.  On the other hand, molecular gas can also
be seen in absorption when back-lighted by strong background
quasars. Using this technique, spectral-line VLBI provides the
possibility to observe molecules even in distant galaxies.  The
canonical example of a molecular absorption system is found toward PKS
1830$-$211, a lensed radio-loud quasar at $z = 2.5$.  A spiral galaxy
at redshift $z = 0.886$ lies along the line of sight, allowing
molecules to be detected at this intermediate redshift (Wiklind \&
Combes 1996).  Muller et al.\ (2011) detected 28 different molecules
and 8 isotopologues in this system (Figure~\ref{fig-absorbers}), which
they note is the extragalactic source with the largest number of
detected molecular species, making it possible to undertake detailed
studies of molecular and isotopic abundance ratios in a source at a
cosmological redshift.

\begin{figure}
\resizebox{\hsize}{!}{\includegraphics{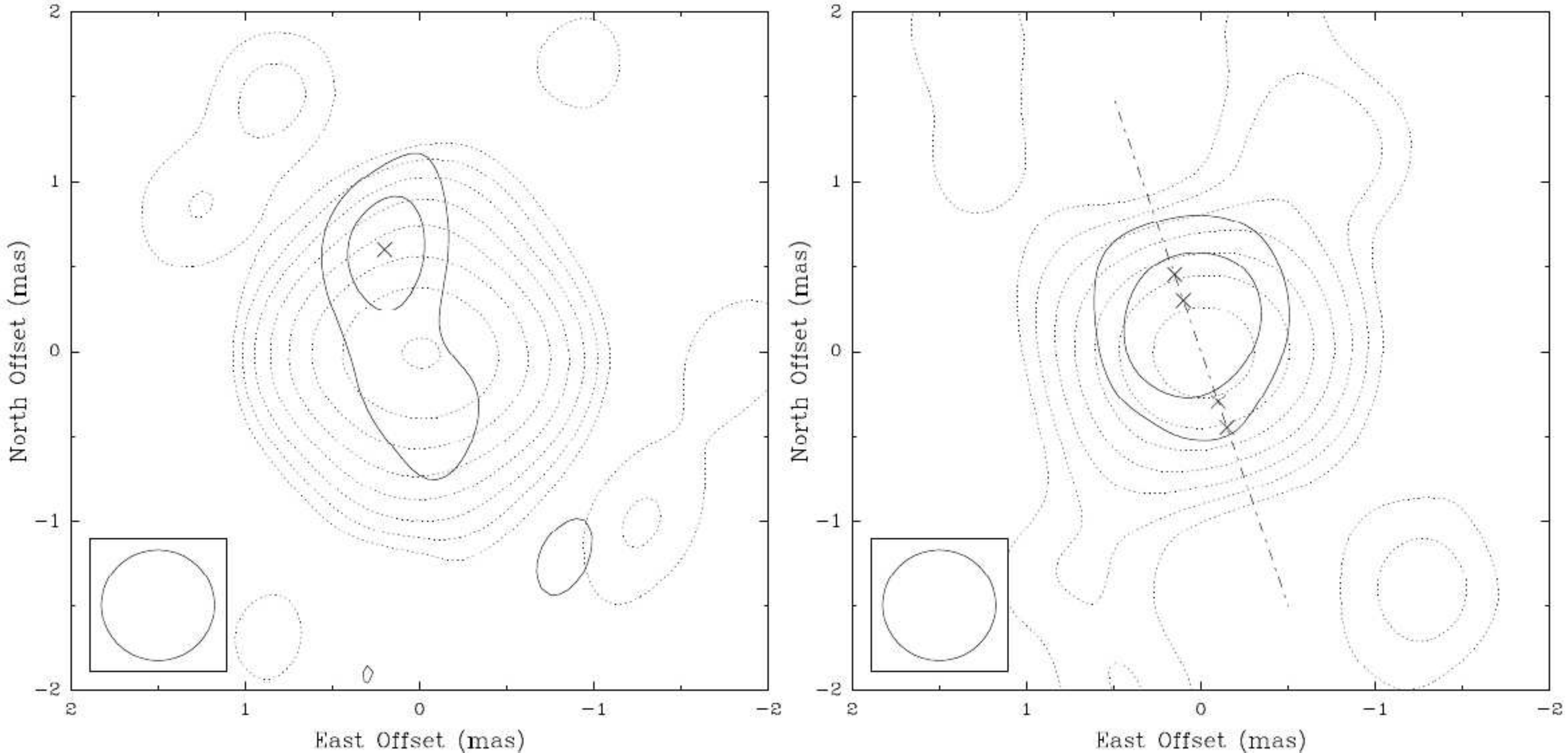}}
\caption{There is structure in the submilliarcsecond distibutions of
  the $J = 3 \leftarrow 2$ (left, solid contours; dotted contours show
  the continuum emission from the background quasar) and $J = 5
  \leftarrow 4$ (right) of HC$_3$N in PKS 1830$-$211 that is not fully
  resolved by the observing beam at centimeter wavelengths (Sato et
  al.\ 2013).  High-frequency VLBI with ALMA will improve
  substantially on the resolution available with current VLBI
  networks.
\label{fig-sato}}
\end{figure}

High angular resolution is important in order to reduce the volume of
gas being observed, which can blend multiple subregions with differing
velocities, abundances, and excitation values.  The spatial
distribution on milliarcsecond scales of cosmologically distant
absorbers has been determined in a couple of systems (e.g., Carilli et
al.\ 2000).  In the case of PKS 1830$-$211, monitoring of the HCO$^+$
and HCN line profiles reveals that the clouds producing the absorption
have sizes smaller than 1~pc (Muller \& Gu\'{e}lin 2008).  Present
VLBI observations at 43~GHz do not resolve the source, although Jin et
al.\ (2003) estimate a size of $1.8 \times 1.2$~pc for the southwest
lensed image after deconvolution of their observing beam.
High-resolution observations of HC$_3$N in this system show that the
spatial distribution of two different transitions on these scales is
different, with the transitions being offset from both the continuum
peak and each other (Figure~\ref{fig-sato}; Sato et al.\ 2013).  The
implied rotational temperature is lower than the expected cosmic
microwave background (CMB) temperature at the absorber's redshift,
which may be due to unabsorbed continuum emission within the observing
beam.  By observing multi-transitions from multiple species in the
millimeter band, Muller et al.\ (2013) could derive a precise and
accurate measurement of the CMB temperature toward
PKS\,1830$-$211. They found a value in agreement with the expected
value at the redshift of the absorber, consistent with adiabatic
expansion of the Universe. Similar observations of other absorber
sources could explore the evolution of the CMB temperature with
redshift, shedding light on the nature of dark energy.

Molecular absorption systems can also be used to place constraints on
the cosmological variations of fundamental constants of nature, like
the fine structure constant ($\alpha$) or the proton-to-electron mass
ratio ($\mu$), as an important test of fundamental physics (e.g., Uzan
2011).  The rest frequency of some spectral lines have different
dependences on fundamental constants.  If these constants vary with
space and/or time (in contradiction to the invariance principle), then
the frequency of such lines will appear to drift and offset in
velocity with respect to other reference transitions. Since kinematic
Doppler shifts can mimic the apparent frequency shifts that would be
produced by variations in fundamental constants, it is preferable to
observe lines that arise from cospatial species, or from a single
suitable species like methanol (Bagdonaite et al.\ 2013).

VLBI with ALMA in Band~6 will improve on the angular resolution in
this system by a factor of more than 30 in solid angle, likely
substantially reducing velocity scatter between molecular transitions
due to contributions from different clouds in the absorbing galaxy.
This will be vitally important for improving estimates of the
deviation of $\mu$.  Henkel et al.\ (2009) measure $|\Delta \mu/\mu| =
(+0.08 \pm 0.47) \times 10^{-6}$ from observations of a range of
different transitions of NH$_3$ and HC$_3$N, but the rms velocity
variation within transitions of each species is nearly 1~km\,s$^{-1}$,
producing a 0.5~km\,s$^{-1}$ random uncertainty in the average
velocities of the two species.  Muller et al.\ (2011) found a
comparable scatter in the centroid velocities of their large number of
different species.  Therefore, it is essential to obtain sensitive
VLBI observations to reduce the systematic uncertainty in $\Delta
\mu/\mu$ due to kinematic and chemical segregation in the absorbing
clouds.  Arrays with baseline lengths on G$\lambda$ scales, such as
millimeter VLBI arrays including ALMA, will be able to resolve these
clouds and even smaller structures, reducing systematic errors in
fundamental constant variation tests.  The sensitivity provided by
phased ALMA may allow the extension of this technique to
as-yet-undetected weaker absorber systems.

The detection of rare isotopologues in distant galaxies also offer the
possibility to investigate the nucleosynthesis history in the Universe
from the evolution of isotopic ratios.  For example, Muller et
al.\ (2006, 2011) could estimate the isotopic abundance ratios of
carbon, nitrogen, oxygen, sulfur and silicon at $z$=0.89 toward PKS
1830$-$211, at a time roughly half the current age of the Universe.
They found significant differences compared to the same isotopic
ratios in the local Universe, and interpreted them as due to a gradual
nucleosynthesis enrichment by low-mass stars, while massive stars
contribute on a much smaller timescale.  With high-angular resolution
and high sensitivity, offered by the VLBI network including ALMA, it
would be possible to investigate the interstellar mixing of elements
at the sub-parsec scale in distant galaxies, and to much smaller
scales by studying molecular absorption in Milky Way clouds.
 
Finally, recent ALMA Early Science data on PKS 1830$-$211 focusing on
absorption lines in the intervening galaxy have revealed remarkable
time- and frequency-dependent variations in the flux ratios between
the two lensed images of the background blazar (Mart\'i-Vidal et
al.\ 2013).  Those variations were interpreted as chromatic structure
changes related to the activity close to the base of the blazar jet,
as the ALMA observations were serendipitously coincident with a strong
gamma-ray flare in the blazar.  Hence, the observations of molecular
absorbers have multiple interests, from the study of absorbing gas in
the foreground galaxy, using molecules as cosmological probes, to the
study of the background (most likely strongly lensed) quasar, vitally
requiring the high angular resolution and high sensitivity that ALMA
will bring to VLBI.

\section{Masers}\label{masers}

Thanks to large amplification factors, many masers are very spatially
compact.  Multi-epoch VLBI studies of masers have been very
scientifically productive, allowing astronomers to understand the
environments and detailed dynamics of star-forming regions (SFRs) and
the envelopes of evolved stars as well as obtain robust geometric
distance measurements of other galaxies and of SFRs in the spiral arms
of the Milky Way.  The main emphasis has traditionally been placed on
observations at centimeter wavelengths (OH and 1.35~cm H$_2$O), but
there has been growing interest in millimeter and submillimeter maser
studies as sensitive high-frequency telescopes and arrays have come on
line (e.g., Humphreys 2007).  These bands provide access to additional
transitions of masing species commonly observed at centimeter and
millimeter wavelengths (e.g., SiO, H$_2$O), as well as new maser
transitions that have been less well studied, particularly with high
spatial resolution (e.g., HCN).  Two major species observed in both
SFRs and circumstellar environments, H$_2$O and SiO, exhibit masing
transitions in ALMA Bands 3 and 5--9, with other observed or predicted
maser transitions existing in most of the remaining ALMA Bands.

\subsection{Galactic Science}

The spectrum of silicon monoxide and its isotopologues includes
transitions at multiples of approximately 43~GHz, many of which are
observed to produce masers in stellar envelopes. The 43 and 86~GHz
transitions of SiO are routinely observed using VLBI techniques, and
the millimeter and submillimeter bands provide access to a number of
high rotational and vibrational SiO transitions detectable in evolved
stars (e.g., Wittkowski et al.\ 2011).  For example, successful VLBI
observations of SiO masers at 129~GHz have been performed in the
envelope of the supergiant VY~Canis Majoris (Doeleman et al.\ 2002),
and multiple SiO transitions have been imaged in a single object
(e.g., Soria-Ruiz et al.\ 2004).

In general, different maser transitions in a molecules and its
isotopologues probe different physical conditions and thus different
regions within a given source.  The simultaneous observation of
multiple transitions therefore enables mapping of the gas kinematics
as well as changes in parameters such as density and temperature as a
function of radius.  The high sensitivity of phased ALMA, together
with its rapid band switching and subarraying, will allow much more
acuurate registration of maser maps in different transitions, as well
as the study of the origin of emission that, present in single-dish
data, is often not detected in current high-resolution VLBI maps.  In
the envelopes of AGB stars, detailed comparison of the ring-like
distributions of different 43 and 86~GHz SiO maser transitions will
provide crucial tests for constraining models of maser pumping
mechanisms, as seen already in Desmurs et al.\ (2000), Soria-Ruiz et
al.\ (2004, 2007), Gray et al.\ (2009), and in Doeleman et al.\ (2004)
for the case of the Orion outflow.

Spatially resolved observations of high-frequency masers will also
provide crucial information about magnetic fields in star-forming
regions and evolved stars (P\'{e}rez-S\'{a}nchez \& Vlemmings 2013).
In particular, highly polarized ($> 20\%$) high-frequency SiO masers
are good probes of the magnetic field morphology within a few stellar
radii of stellar atmospheres (e.g., Vlemmings et al.\ 2011).  While
ALMA by itself has insufficient resolution, VLBI including phased ALMA
will be able to map in detail the magnetic field on very small scales.
Multi-epoch observations could trace magnetic ejections from evolved
stars.  Although current magnetic field studies have focused on
oxygen-rich stars, the 89~GHz HCN maser can be used for similar
studies of carbon-rich envelopes.

In addition to magnetic field information, maser polarization provides
unique constraints on maser theory models when comparing different
transitions.  Previous work has been limited to the 43 and 86~GHz SiO
masers, and only one source has been studied in detail (Richter et
al.\ 2012).  With sensitive millimeter VLBI, many more sources and maser
transitions will be available for study, and it will finally be
possible to investigate predicted correlations between linear and
circular polarization fractions as well as between linear polarization
fractions of maser features at different frequencies.

While SiO masers are known in hundreds of late-type stars, they have
been detected in only 3 high-mass SFRs: Sgr~B2, W51, and Orion.  SiO
masers have been extensively mapped in the direction of the Orion
compact source I.  At centimeter wavelengths, proper motion
measurements of these masers provide a unique probe of accretion and
outflow processes within $\sim 10$-100~AU of the source (e.g.,
Matthews et al.\ 2010). Application of millimeter VLBI to the study of
this source would enable more complete mapping of the material through
access to numerous additional SiO transitions. The previous
observations at 43~GHz with $\sim 500~\mu$as resolution by Matthews et
al.\ show that, in contrast with what is observed in late-type stars,
nearly all the flux is recovered at that resolution, suggesting that
the type of sensitive, higher resolution studies enabled by millimeter
VLBI with phased ALMA are likely to be fruitful for improving proper
motion studies and providing more detailed mapping of the gas
kinematics.  As in the case of evolved stars, simultaneous
measurements of multiple SiO transitions surrounding Source~I will
also provide tests of maser excitation mechanisms (e.g., Goddi et
al.\ 2009).  Initial findings by Matthews et al.\ suggest these may be
quite different from the case for evolved stars.

Submillimeter water masers expected in ALMA Bands 7--9 are also
relevant to studies of star formation.  Besides the ubiquitous 22~GHz
line, strong water lines have been detected at 183, 321, and 325~GHz
and higher frequencies in SFRs as well as in several circumstellar
envelopes of late-type stars using single-dish telescopes (Humphreys
et al.\ 2007 and references therein).  Their spectra show
spectacularly intense peak fluxes (up to 10000~Jy) and large velocity
ranges (up to 300~km\,s$^{-1}$).  Neufeld \& Melnick (1991) and more
recently Daniel \& Cernicharo (2013) discussed collisional pumping for
these masers and predicted simultaneous inversion of 22, 321, and
325~GHz and other submillimeter H$_2$O transitions over a wide range
of physical conditions.  Recent ALMA science verification observations
have produced the first detection of high-excitation 232.686~GHz water
masers associated with Orion~KL Source~I (Hirota et al.\ 2012), a
vibrationally excited masing transition that, like SiO masers, was
known to be excited in late-type stars; this is the first detection in
a SFR.  In principle, observations of different maser line ratios may
be used to constrain physical conditions; in practice, this has not
been possible so far because single-dish antennas and
connected-element interferometers can only describe the average
emission of many maser clumps.  VLBI images of submillimeter H$_2$O
maser lines are required to provide high enough angular resolution to
resolve these clumps, constraining and testing radiative transfer
models and mapping gas temperature and density with high accuracy and
resolution around young stellar objects.

So far, only two SFRs, Orion~KL and Cep~A, have been imaged in
submillimeter H$_2$O lines with the SMA.  Challenging weather
requirements (especially at 325~GHz) and poor dynamic range have
prevented imaging of other SFRs, limiting their use as probes of
high-mass star formation.  ALMA's superb high site meets weather
requirements for observations at 321 and 325~GHz as well as for even
higher-frequency transitions of interest in SFRs and late-type stars
(e.g., 439 and 471~GHz).  Inclusion of ALMA in submillimeter VLBI
arrays will enable studies of submillimeter H$_2$O maser lines in a
larger sample of SFRs and permit comparison with centimeter lines
observed at submilliarcsecond resolution.  Polarimetric observations
of maser lines can also provide measurements of magnetic fields in
outflows in SFR outflows (H$_2$O masers) and in circumstellar
envelopes (SiO, HCN, and H$_2$O).

Lastly, VLBI with phased ALMA could open an entire new domain for the
study of the envelopes of carbon stars, where HCN masers are believed
to play a role analogous to those of SiO masers in oxygen-rich stars
(e.g., Izumiura et al.\ 1995).  This has remained until now a
relatively unexplored area owing to the lack of availability of
sensitive millimeter and submillimeter VLBI facilities.

\subsection{Extragalactic Science}

Luminous extragalactic masers, namely the ``megamasers,'' are mainly
produced by two molecules: OH and H$_2$O.  While OH megamasers are in
most cases related to particularly high star formation in
(ultra)luminous infrared galaxies, water megamasers are associated
with AGN actitivy, either with accretion disks, nuclear outflows, or
radio jets (e.g., Lo 2005; Tarchi 2012).

Extragalactic water masers are very compact and luminous.  The
canonical example of a circumnuclear water maser source, NGC 4258,
contains 1.35~cm water masers embedded in a warped accretion disk
around a supermassive black hole (Miyoshi et al.\ 1995).  The
Keplerian rotation of this disk as traced by the masers allows a
purely geometric distance to be obtained (Herrnstein et al.\ 1999).
The phenomenal success of modeling NGC 4258 is the inspiration for the
Megamaser Cosmology Project (e.g., Reid et al.\ 2009b, 2013; Kuo et
al.\ 2011), which has identified over 100 galaxies hosting megamasers
to which this technique can be applied in order to measure the Hubble
constant directly.

While the main focus has been on the 22~GHz transition that is easily
accessible with existing VLBI arrays, millimeter and submillimeter
water masers are also being targeted.  These higher-frequency masers
may allow the back side of the circumnuclear disk to be observed,
since the disk is optically thick at 22~GHz; would provide constraints
on radiative transfer models, from which the density and temperature
of the disks could be inferred; and would be observable with higher
angular resolution, which is important for extending the technique to
more distant galaxies.  One promising source is NGC 3079, a nearby
Seyfert 2/LINER galaxy in which water masers have been detected at 183
and 439~GHz (Humphreys et al.\ 2005).  Modelling of the 325 GHz~line
suggests that it can be inverted both under the same conditions that
will produce 22~GHz masers and at lower densities and temperatures
(e.g., Cernicharo et al.\ 2006), potentially making it an excellent
probe of both the circumnuclear disk and the wide-angle outflow that
may be associated with the large-scale superbubble.  Cernicharo et
al.\ (2006) find 183~GHz water masers in Arp~220---a galaxy that hosts
OH megamasers but no 22~GHz water masers---, which is an intriguing
target for further study as water and OH megamasers are only very
infrequently found in the same galaxy (Tarchi et al.\ 2011).

Very luminous Galactic-analogue water kilomasers are also seen in
extragalactic star-forming regions (e.g., Tarchi et al.\ 2002;
Hagiwara 2007; Surcis et al.\ 2009).  However, some kilomasers may be
related to AGN activity as well (Tarchi et al.\ 2011).  VLBI
observations of submillimeter water masers may provide an extremely
high angular resolution picture of star formation in nearby galaxies
as well as new insights on the putative class of nuclear kilomasers.

\section{Astrometry}\label{astrometry}

Precision astrometric observations of target sources require phase
connection with a nearby calibrator.  The ideal case occurs when both
the target and the calibrator source can be observed within the
primary beam of each VLBI element.  A more typical observing mode is
to nod between the target and calibrator sources on a timescale faster
than the atmospheric coherence time.  This strategy can be employed in
Band 1 and, under good weather conditions across the observing array,
Band 3.  At higher frequencies, where the atmospheric coherence time
may be too short for nodding, astrometry could still be accomplished
by observing with phased-array stations that are subarrayed to observe
the target and calibrator sources simultaneously.

\begin{figure}
\resizebox{\hsize}{!}{\includegraphics{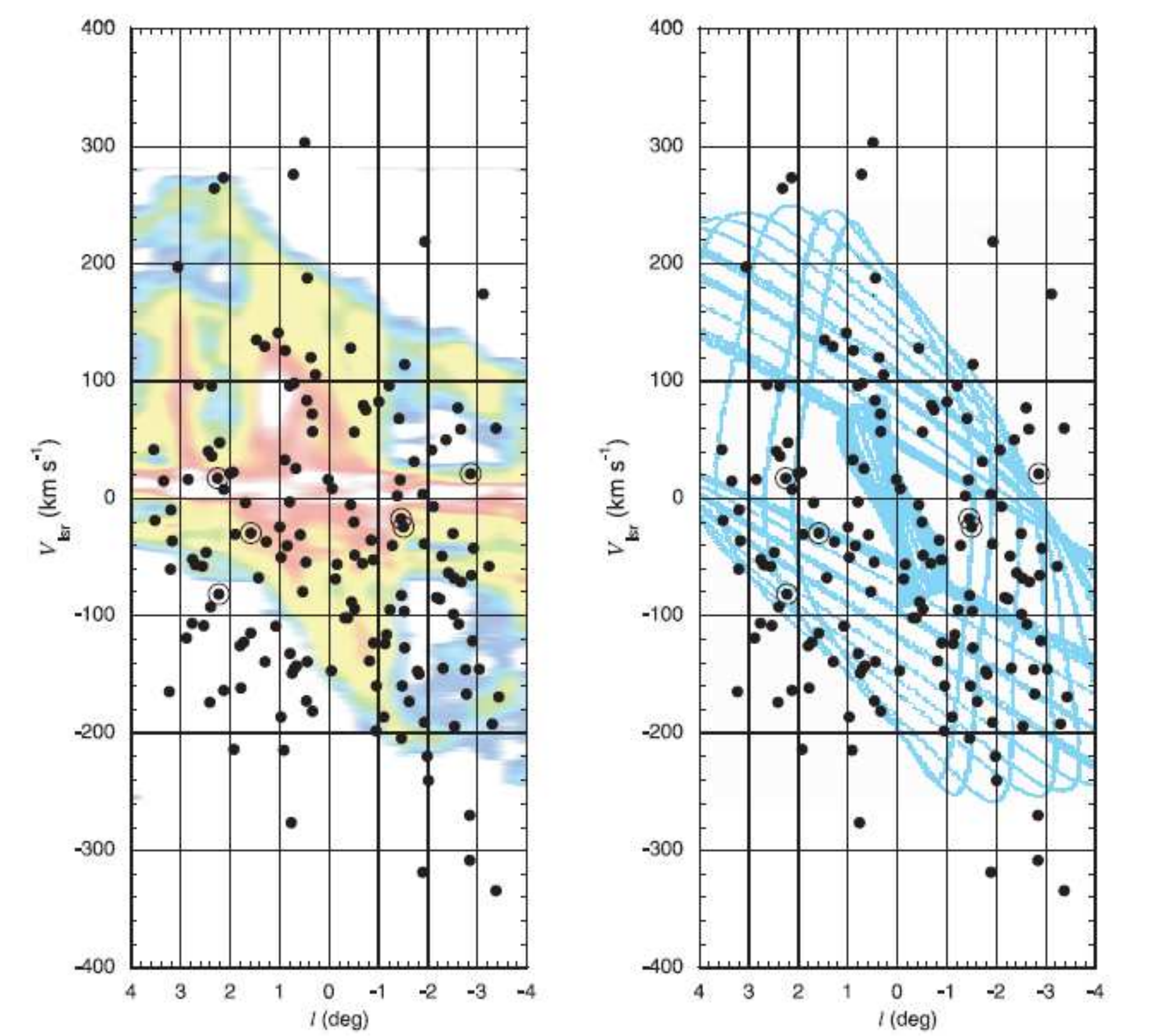}}
\caption{Left: Longitude-velocity diagram of SiO maser sources in the
  Galactic bulge, superposed atop CO line emission (Fujii et
  al.\ 2006).  Right: Stellar orbits in the bar potential.
\label{fig-fujii}
}
\end{figure}

Astrometry is complementary to multiple aspects of jet science
(\S~\ref{agn}).  Multifrequency astrometric observations played a
crucial role in identifying the location of the black hole in M87
(Hada et al.\ 2011) and could be extended to higher frequencies to
determine the location of the 1.3~mm emission, which is associated
with the inner accretion and outflow region (Doeleman et al.\ 2012),
relative to longer-wavelength observations of the forward jet and
counterjet.  This would enable more accurate measurements of the jet
opening angle, inclination to the line of sight, and Lorentz factor
(Broderick et al.\ 2011c).

Sgr A* is known to be variable across the spectrum on timescale of
minutes to hours (e.g., Genzel et al.\ 2003; Eckart et al.\ 2006;
Marrone et al.\ 2008), although it is not known which of several
mechanisms may be responsible for this variability (e.g., orbiting hot
spots, expansion of material ejected from the disk, or episodic
ejection of material along a jet).  A VLBA search for the position
wander of Sgr~A* at 7~mm was able to place constraints on, for
instance, the amount of the total source brightness that could be
contained within an orbiting hot spot at fairly large radius
($15~GM/c^2$; Reid et al.\ 2008), but the inclusion of phased ALMA in
astrometric arrays could improve on this substantially (Broderick et
al.\ 2011c).  The proper motion of Sgr~A* has also been measured with
the VLBA at 7~mm (Reid \& Brunthaler 2004), but the astrometric
improvements provided by the inclusion of phased ALMA in millimeter
observing arrays would permit the measurement of microarcsecond-scale
fluctuating components induced by stars and stellar-mass black holes
in the Galactic Center (Broderick et al.\ 2011c).  Absolute
astrometric observations of Sgr~A* will also directly provide the
distance to the Galactic Center, $r_0$, which is one of the most
fundamental parameters in the field of Galactic astronomy.
Lower-frequency Northern hemisphere VLBI arrays have attempted to
measure the parallax of Sgr~A*, but prospects of success have been
limited by severe interstellar scattering.  High-frequency
observations including ALMA and other stations, especially in the
Southern hemisphere, will be best-suited for absolute astrometric
measurements of Sgr~A*.

Astrometric observations of maser sources have been very productive in
establishing geometric distances to Galactic objects and nearby
extragalactic sources.  For instance, centimeter-wavelength
observations as part of the Bar and Spiral Structure Legacy (BeSSeL)
Survey and VLBI Exploration of Radio Astrometry (VERA) have
substantially revised the IAU recommended values for the Solar motion
and Galactocentric Solar distance and are beginning to clarify the
nearby spiral structure of the Milky Way (Brunthaler et al.\ 2012;
Honma et al.\ 2012).  Astrometric observations in Bands 1 and 3 of the
abundant SiO masers in the Galactic bar and nuclear region similarly
have the potential to be of great help in understanding the dynamics
of the inner Galaxy.  In fact, there are hundreds of SiO maser
sources---mostly supergiant stars---found toward the Galactic bulge
and Galactic Center regions.  Figure~\ref{fig-fujii} shows the
distribution of SiO maser sources toward the Galactic Center (Fujii et
al.\ 2006).  Some stars have a high velocity and even a forbidden
velocity in terms of gas motion.  This is most likely due to the bar
potential; since the stars are collisionless and the gases
collisional, their motions in the bar potential are expected to be
different.  Hence, astrometry of SiO maser sources in the bulge is
complementary to astrometry with BeSSeL and VERA, which have been
observing star-forming region, and will test predictions of Galactic
bar models based on the three-dimensional spatial motions of stars and
gas clouds.  SiO maser astrometry also provides a unique opportunity
to cross-register radio and infrared reference frames at
submilliarcsecond accuracy (e.g., Menten et al. 1997).  In addition,
absolute astrometry of SiO masers in late-type stars will allow
accurate alignment of maps of various maser transitions with each
other, constraining the pumping mechanisms at work, and with the
underlying optical star, as more accurate absolute positions are
obtained with the Gaia space observatory.

Southern hemisphere VLBI arrays including phased ALMA will also be
powerful for astrometry of the portion of the Galaxy inaccessible to
northern arrays such as the VLBA and VERA.  SiO masers in supergiant
stars are again the most promising targets to reveal the structure and
dynamics of the spiral structure in the third and fourth Galactic
quadrants.  An unbiased pilot survey of SiO masers in the southern
Galactic plane using the Mopra 22-m telescope indicates that numerous
suitable SiO maser target sources exist in this region (e.g., Jordan
et al.\ 2013).

Looking beyond the Milky Way, these techniques could be used in
conjunction with other VLBI-capable telescopes in the Southern
Hemisphere to obtain a direct geometric distance to SiO masers in the
Large Magellanic Cloud (LMC; van Loon et al.\ 2001), potentially
improving upon Cepheid distances, which contain systematic
uncertainties due to reddening and metallicity (An et al.\ 2007).
Since SiO maser emission in the LMC is faint due to its distance, such
observations cannot be done without the sensitivity provided by phased
ALMA.  Another application of high-precision astrometry is to measure
the motions of nearby galaxies with radio AGN.  A source motion of
500~km\,s$^{-1}$ at a distance of 10~Mpc produces a proper motion of
$\sim 10~\mu$as\,yr$^{-1}$.  Proper motion measurements of sources in
the Virgo cluster ($\sim 16$~Mpc) could be done within a few years,
providing information on cluster dynamics.

Selected compact extragalactic radio sources serve as fiducial points
to define the celestial reference frame.  Based on several
decades-long astrometric VLBI measurement time series, many of them
show apparent proper motions of $\sim$0.01--0.1~mas\,yr$^{-1}$.  It is
crucial to understand the origin of these proper motions, to define a
more accurate reference frame, and to accurately study phenomena like
the secular aberration drift caused by the acceleration of the solar
system barycenter due to the rotation around the Galactic center
(e.g. Titov et al.\ 2011).  While the characteristic direction of
VLBI-measured AGN proper motions seems generally connected with the
$\sim$1-10-mas scale jet structure, there are cases of significant
misalignment (Mo\'or et al.\ 2011).  In particular, the proper motion
direction for OJ~287 is nearly orthogonal to its radio jet seen with
centimeter-wavelength VLBI.  The relationship between the small
apparent proper motions and the brightness structure could be
established with sensitive millimeter-VLBI imaging, which probes the
right angular scales, well within 1~mas from the central engine.

\section{ALMA Interferometric Datasets}

It is important to note that the standard operating mode for the ALMA
beamformer will include archiving of the products of the ALMA Baseline
Correlator.  Observations with the ALMA beamformer therefore will
additionally provide ``regular'' ALMA interferometric datasets.  VLBI
campaigns could end up producing deep ALMA integrations on certain
targets.  For example, the ALMA interferometric dataset corresponding
to a series of 230~GHz VLBI observations totalling 30 hours on source
would have an rms sensitivity of $\sim 2$~$\mu$Jy\,beam$^{-1}$ for
continuum sources and 0.2~mJy\,beam$^{-1}$ for a single
1.3~km\,s$^{-1}$ channel in the standard widest-bandwidth observing
mode.

One field that will likely see high interest from VLBI proposers is
the Galactic Center region.  At 230~GHz, the ALMA primary beam of
26~arcsec covers the central parsec of the Galaxy, which includes
Sgr~A*, most of the minispiral, and a portion of the circumnuclear
disk.  Deep continuum sensitivities will allow spectacular mapping of
the morphology and polarization of the continuum emission, and
spectral indices and rotation measure estimates can be determined from
comparison of the measurements between the two sidebands.  Variability
can be probed from timescales much smaller than a single VLBI epoch to
multi-epoch or even multi-year timescales, and polarimetric
variability can be used to constrain changes in the Faraday rotation
screen toward the Galactic Center.  Flexibility with VLBI tuning at
ALMA and other facilities would also permit inclusion of particular
spectral lines of interest.

Deep-field ALMA datasets may be obtained in the direction of other
targets based on their VLBI scientific potential.  Depending on the
pointing direction, these observations may contain sources in the
Galactic plane or halo, nearby galaxies, and/or high-redshift
galaxies.  Examples of the scientific potential of these ALMA
interferometer datasets can be found in the ALMA Design Reference
Science
Plan\footnote{\url{http://www.eso.org/sci/facilities/alma/documents/drsp.html}}.

\section{Concluding Remarks}

The preceding sections summarize areas in which the ALMA beamformer
will be uniquely capable to have great scientific impact.  These span
a wide range: from astrophysics to fundamental physics, from Galactic
sources to objects at cosmic redshift, from very compact objects to
kiloparsec-scale jets, and from cool molecules to extremely hot
plasmas.  Most of these studies will be made possible thanks to the
sensitivity and angular resolution that phased ALMA will bring to VLBI
arrays.  Pulsar observations with the ALMA beamformer can be done
commensally with standard ALMA observing too, increasing the
scientific efficiency of allocated ALMA time.

In this document we have attempted to identify the most fruitful lines
of investigation that will be enabled by the ALMA beamformer.
However, the history of astronomy demonstrates that when an instrument
opens up a new area of discovery space, some of the most important
scientific results come from unanticipated directions.  The full
measure of the scientific capability of the phased ALMA system will
ultimately be limited only by the imagination of the astronomy
community.

\acknowledgments

This work is funded by the National Science Foundation (MRI AST-1126433).

\pagebreak
\clearpage

\mbox{}
\end{document}